\documentclass{article}

\PassOptionsToPackage{numbers, compress}{natbib}

\usepackage[pagebackref,breaklinks,colorlinks]{hyperref}

    \usepackage[preprint]{neurips_2023}
    \usepackage{pdfpages}

\usepackage[utf8]{inputenc} 
\usepackage[T1]{fontenc}    
\usepackage{hyperref}       
\usepackage{url}            
\usepackage{booktabs}       
\usepackage{amsfonts}       
\usepackage{nicefrac}       
\usepackage{microtype}      
\usepackage{xcolor}         
\usepackage{tabularray}
\usepackage{caption}
\usepackage{soul}
\usepackage{graphicx}
\usepackage{float}
\usepackage{changepage}
\usepackage{adjustbox}
\usepackage{microtype}

\title{Pegasus-1 Technical Report}

\author{%
   \\
}

\begin{document}

\maketitle

\newcommand{\todoc}[2]{{\textcolor{#1}{\textbf{#2}}}}
\newcommand{\todoblue}[1]{\todoc{blue}{\textbf{#1}}}
\newcommand{\todored}[1]{\todoc{red}{\textbf{#1}}}
\newcommand{\todogreen}[1]{\todoc{green}{\textbf{#1}}}
\newcommand{\todoorange}[1]{\todoc{orange}{\textbf{#1}}}
\newcommand{\todomagenta}[1]{\todoc{magenta}{\textbf{#1}}}

\newcommand{\hyojun}[1]{\todoblue{\textbf{hyojun:} #1}}
\newcommand{\flynn}[1]{\todogreen{\textbf{flynn:} #1}}
\newcommand{\ray}[1]{\todored{\textbf{ray:} #1}}
\newcommand{\pegasus}{Pegasus-1~}
\newcommand{\minjoon}[1]{\todoorange{\textbf{mj:} #1}}
\newcommand{\jay}[1]{\todomagenta{\textbf{jay:} #1}}

\begin{figure}[ht]
\centering
\vspace{-50pt}
\includegraphics[width=0.4\linewidth]{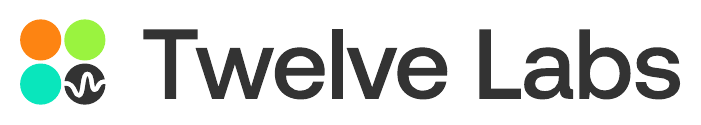} 
\label{fig:yourlabel} 
\end{figure}
\vspace{20pt}

\begin{abstract}
This technical report introduces Pegasus-1, a multimodal language model specialized in video content understanding and interaction through natural language.\footnote{Please cite this paper as (Twelve Labs, 2024). Please see Authorship Section at the end of this report for the full list of contributors.}
Pegasus-1 is designed to address the unique challenges posed by video data, such as interpreting spatiotemporal information, to offer nuanced video content comprehension across various lengths.
This technical report overviews Pegasus-1's architecture, training strategies, and its performance in benchmarks on video conversation, zero-shot video question answering, and video summarization.
We also explore qualitative characteristics of Pegasus-1 , demonstrating its capabilities as well as its limitations, in order to provide readers a balanced view of its current state and its future direction.\footnote{Pegasus-1 blog post and API are available at  \url{https://www.twelvelabs.io/blog/upgrading-pegasus-1}}
\end{abstract}

\section{Introduction}
In the evolving landscape of large language models (LLMs)~\cite{achiam2023gpt, touvron2023llama, chiang2023vicuna, jiang2023mistral, chowdhery2023palm}, cultivating the video understanding capabilities of LLMs~\cite{li2023mvbench, zhang2023video, zhang2023llama, li2023videochat, maaz2023video, liu2023one, li2023llama,reid2024gemini,team2023gemini} has emerged as a frontier of innovation and practical utility.
This technical report introduces and analyzes Pegasus-1, a state-of-the-art multimodal model that offers versatile capabilities in interpreting, generating, and interacting with video content through natural language.

The major goal of developing Pegasus-1 is to overcome inherent challenges in video data that contain multiple modalities within a single format.
Critical to this understanding is to interpret the temporal sequence of visual data, capturing the essence of movement and change over time while providing a spatially detailed analysis within each frame of the video.
Concurrently, audio information should be incorporated to enhance the interpretation of visual elements and ensure a nuanced understanding of video content.
Moreover, handling a wide range of video lengths is essential to cover various video types from short clips to extended footage.

In this technical report, we discuss our approach to dealing with these challenges to enable Pegasus-1 to understand video content comprehensively. 
The discussion will include a brief description of its model architecture, training data, and training strategies, providing a broad perspective on the factors contributing to Pegasus-1's advanced video understanding capabilities.

Pegasus-1 achieves new state-of-the-art results in video conversation benchmark~\cite{li2023videochat}, zero-shot video question answering~\cite{xiao2021next,yu2019activitynet}, and video summarization~\cite{pegasus-1,pegasus-1-beta}. By outperforming both open-source and proprietary models, Pegasus-1 demonstrates its generalization capabilities in understanding complex video data.

This report also presents a wide range of qualitative results to offer insights into Pegasus-1's capabilities and to explore potential emerging use cases.
Our objective is to unveil and preview the range of functionalities Pegasus-1 can potentially deliver, acknowledging that while these capabilities showcase new possibilities, they may not yet be fully reliable or consistent.
Through this exploration, we aim to highlight Pegasus-1's potential to catalyze new applications and advancements in the field.

Despite its capabilities, Pegasus-1 has inherent limitations that users should be aware of, as we aim for transparency while continuing to refine and enhance its functionalities.
These limitations are briefly discussed within this technical report.
Our objective is to provide users with a comprehensive understanding of Pegasus-1's current strengths, weaknesses and areas for growth, ensuring informed usage and setting accurate expectations for its performance and applicability.

\section{Model Architecture and Training}

\subsection{Model Architecture}

\begin{figure}[t]
    \centering
    \includegraphics[width=\textwidth]{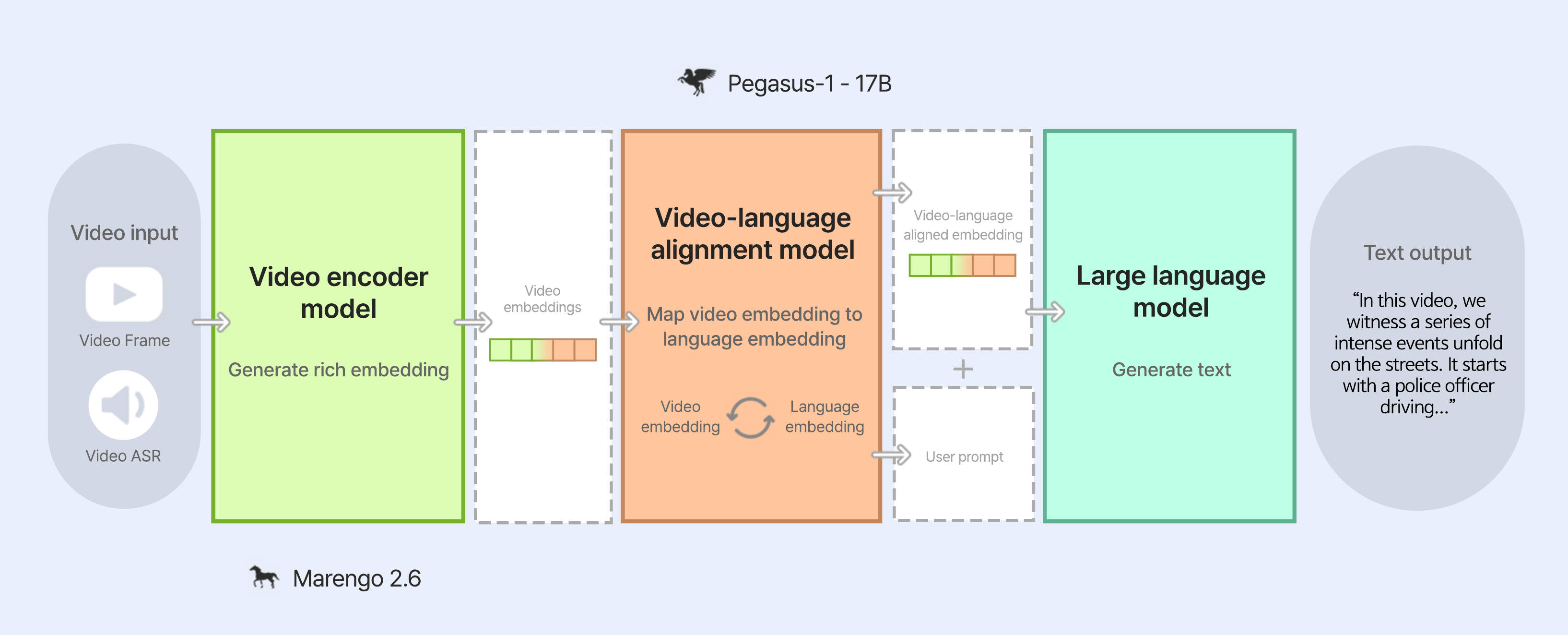}
    \caption{Architectural Overview of Pegasus-1: Pegasus-1 is comprised of three main components: 1) the Video Encoder model for generating multi-modal embeddings from visual and audio inputs, the Video-Language Alignment Model for synchronizing video and text representations, and the Large Language Model for generating contextually relevant textual output.}
    \label{fig:pegasus-arcthiecture}
\end{figure}


To efficiently comprehend video data, it is crucial to seamlessly integrate both auditory and visual information while effectively managing longer video lengths. In our architectural strategy, we have developed an architecture capable of jointly processing audio and visual data, and it is distinguished by its efficient computation tailored for longer videos. As depicted in Fig.~\ref{fig:pegasus-arcthiecture}, Pegasus-1 is structured into a comprehensive tripartite framework designed to encode, align, and decode:
\begin{itemize}
    \item \textbf{Video Encoder Model:} Powered by our Marengo 2.6~\cite{marengo-2.6}, the video encoder model processes the video input to produce rich embeddings from both video frames and audio speech recognition (ASR) data. These embeddings are dense representations that capture the visual and auditory essence of the video content. 

    \item \textbf{Video-language Alignment Model:} The model maps video embeddings to corresponding language embeddings, establishing a shared space where video and text representations are coherently aligned. Such an alignment is pivotal, enabling the model to discern and correlate the visual content of the video with its descriptive language effectively. Additionally, to enhance the processing of longer video lengths, we employ token reduction techniques aimed at minimizing the input token sizes within the large language model, thereby optimizing computational efficiency and maintaining context integrity over longer durations.
    
    \item \textbf{Large Language Model (Decoder Model):} The large language model decoder takes the aligned embeddings and user prompts to generate coherent and contextually relevant text output. This output can range from descriptive summaries to answers to specific questions about the video content. This large language model decoder leverages efficient attention mechanisms for efficiently handling the long context of video data.
\end{itemize}

\subsection{Training}

The training of Pegasus-1, a state-of-the-art multimodal foundation model, focuses on the harmonious understanding of various modalities in video data, including visual and audio modes. 
To achieve this, we capitalized on the unique strengths of each modality: employing spatial understanding from image data and deriving insights from audio data for robust audio understanding.
Notably, despite the video having a richer information source compared to standalone images or audio, there is a marked scarcity of high-quality video data. 
We address this challenge by amassing an extensive collection of proprietary text annotations for training.
Specifically, we annotate 10M+ diverse videos with highly detailed descriptions. These descriptions capture most of the events that appear in each video, incorporating detailed real-world knowledge about activities and objects, including specific names of objects and technical terms for various sports.\footnote{Note that the number of tokens per video depends on its length and spatiotemporal resolution. For instance, if one assumes 1 frame per second, 256 patches per frame, and 3-minute (180s) per video for simplicity, each video would be equivalent to around 46k tokens and 10M videos would be equivalent to 0.46T tokens.}

Our training process for Pegasus-1 consists of two phases: pretraining phase and instruction tuning phase.
During the pretraining phase, we initialize both the video encoder and large language model with pre-trained weights, subsequently training with our expansive multi-modal datasets, which include the rich detailed video data mentioned previously.
In the second phase, we conduct supervised instruction finetuning, utilizing proprietary multimodal instruction datasets to refine the model's responsiveness to user instructions.
However, a significant challenge in such multi-stage training is the risk of catastrophic forgetting~\cite{mccloskey1989catastrophic}, where the pretrained model loses previously acquired knowledge upon assimilating new information.


To mitigate this, our training strategy incorporates two meticulously planned stages, aiming to harmonize the integration of new data while preserving existing knowledge. 
The cornerstone of this strategy is the selective unfreezing of model parameters combined with precise adjustments to learning rates during the training progression.
This approach ensures that Pegasus-1 not only acquires new capabilities efficiently but also retains and refines its previously learned skills, establishing a robust foundation for advanced video-language understanding.

\section{Results on Benchmarks}

We evaluate Pegasus-1's performance across well-established video large language model benchmarks, including video conversation~\cite{maaz2023video}, zero-shot video question answering~\cite{yu2019activitynet,xiao2021next}, and video summarization~\cite{pegasus-1}.
This section presents the performance of Pegasus-1 in comparison to both proprietary and open-source models.

\subsection{Comparison Models}

We compare our model against the following baselines:
\begin{itemize}
    \item \textbf{Gemini models (proprietary)}~\cite{team2023gemini,reid2024gemini}: Gemini is Google's proprietary multimodal model, also known for its state-of-the-art performance in video-language tasks. To ensure the accuracy and relevance of our comparative analysis, we excerpt the performance of Gemini 1.0 and Gemini 1.5 from Google's technical report for the Video Question Answering benchmark. For other benchmarks that are not included in their reports, such as video conversation and summarization, we run the February 17 2024 version of the publicly available Gemini API.\footnote{Gemini API documentation: ~\url{https://github.com/GoogleCloudPlatform/generative-ai/blob/main/gemini/getting-started/intro_gemini_pro_vision_python.ipynb}} Note that Gemini 1.5 Pro API is not available as of March 30th, 2024 and is only accessible through its playground-like preview access.
    \item \textbf{Open-source models}: We compare Pegasus-1 with VideoChat~\cite{li2023videochat}, Video-ChatGPT~\cite{maaz2023video}, Video LLAMA~\cite{zhang2023video}, BT-adapter~\cite{liu2023one}, LLaMA-VID~\cite{li2023llama}, and VideoChat2~\cite{li2023mvbench}. We use the benchmark results reported in their papers if they are available.
\end{itemize}

\subsection{Results}

\paragraph{Video conversation benchmark.}

\begin{table}
\centering
\begin{tabular}{l|ccccc}
\toprule
\textbf{Model}         & \textbf{Correctness} & \textbf{Detail} & \textbf{Context} & \textbf{Temporal} & \textbf{Consistency} \\
\midrule\midrule
VideoChat~\cite{li2023videochat}     & 2.23        & 2.5    & 2.53    & 1.94     & 2.24        \\
Video-ChatGPT~\cite{maaz2023video} & 2.4         & 2.52   & 2.62    & 1.98     & 2.37        \\
Video LLAMA~\cite{zhang2023video}  & 1.96        & 2.18   & 2.16    & 1.82     & 1.79        \\
BT-adapter~\cite{liu2023one}   & 2.68        & 2.69   & 3.27    & 2.34     & 2.46        \\
LLaMA-VID~\cite{li2023llama}     & 3.07        & 3.05   & 3.6     & 2.58     & 2.63        \\
VideoChat2~\cite{li2023mvbench}    & 3.02        & 2.88   & 3.51    & 2.66     & 2.81        \\
\midrule
Gemini Pro~\cite{team2023gemini}    & 2.98        & 2.99   & 3.44    & 2.32     & 2.32        \\
\midrule
\textbf{Pegasus-1}     & \textbf{3.79}        & \textbf{3.76}   & \textbf{4.29}    & \textbf{3.34}     & \textbf{4.03}     \\   
\bottomrule
\end{tabular}
\vspace{5pt}
\caption{In the comparison conducted using the video conversation benchmark~\cite{maaz2023video}, the open-ended generation outputs are evaluated by a language model (LLM) through a comparison with the ground truth answers. This evaluation specifically focuses on five distinct aspects. Notably, Pegasus-1 outperforms both open-source models and a leading proprietary model. These results were compiled in February 2024. At that time, the Gemini 1.5 API was not yet generally available, hence its exclusion from the comparative table.}
\label{tab:video_conversation_benchmark}
\end{table}

\begin{table}[htbp]
\resizebox{\columnwidth}{!}{%
\begin{tabular}{@{}c|cc|cc|cc|c|cccc|c@{}}
\hline
\multicolumn{1}{l|}{} & \multicolumn{2}{c|}{\textbf{Action}}                 & \multicolumn{2}{c|}{\textbf{Direction}}                & \multicolumn{2}{c|}{\textbf{Speed}}                        & \multicolumn{1}{l|}{\textbf{\begin{tabular}[c]{@{}l@{}}Event\\ Order\end{tabular}}} & \multicolumn{4}{c|}{\textbf{\begin{tabular}[c]{@{}c@{}}Attribute\\ Change\end{tabular}}}                                           & \multicolumn{1}{l}{\textbf{Avg.}} \\ \cline{2-13} 
                      & \multicolumn{1}{c|}{\textbf{Fine}} & \textbf{Coarse} & \multicolumn{1}{c|}{\textbf{Object}} & \textbf{Camera} & \multicolumn{1}{c|}{\textbf{Absolute}} & \textbf{Relative} & \textbf{Order}                                                                      & \multicolumn{1}{c|}{\textbf{Color}} & \multicolumn{1}{c|}{\textbf{Size}} & \multicolumn{1}{c|}{\textbf{Combined}} & \textbf{Other} & \textbf{}                         \\ \cline{2-12}
Random                & 39.7                               & 40.1            & 39.8                                 & 39.0            & 40.8                                   & 42.0              & 41.5                                                                                & 40.4                                & 39.9                               & 38.9                                   & 39.4           & 40.5                              \\
mPLUG-Owl             & 48.8                               & 66.1            & 38.7                                 & 36.8            & 42.2                                   & 38.4              & 42.0                                                                                & 41.7                                & 44.7                               & 41.9                                   & 39.9           & 44.4                              \\
V-LLaVA               & 55.0                               & 83.1            & 41.9                                 & 42.1            & 44.7                                   & 41.6              & 46.2                                                                                & 45.7                                & 45.7                               & 43.5                                   & 49.8           & 49.6                              \\
VideoChat2            & 60.4                               & 77.2            & 45.0                                 & 40.8            & \textbf{52.6}                          & 41.5              & 44.8                                                                                & 50.8                                & 47.9                               & \textbf{49.9}                          & 44.3           & 50.8                              \\ \hline
Pegasus-1             & \textbf{79.8}                      & \textbf{92.6}   & \textbf{46.1}                        & \textbf{44.4}   & 48.1                                   & \textbf{44.2}     & \textbf{56.3}                                                                       & \textbf{51.7}                       & \textbf{49.6}                      & 46.7                                   & \textbf{45.7}  & \textbf{57.1}       \\ \bottomrule
\end{tabular}%
}
\vspace{5pt}
\caption{TempCompass~\cite{liu2024tempcompass} assesses the temporal understanding capabilities of video language models across five dimensions. Pegasus-1 clearly outperforms competing open-source models.}
\label{tab:tempcompass}
\end{table}

In Table~\ref{tab:video_conversation_benchmark}, we present the evaluation results of Pegasus-1 on a video-based conversation benchmark~\cite{maaz2023video}.
The results indicate that Pegasus-1 performs robustly in the context of video conversations, achieving notable performance in various evaluative dimensions.
Specifically, Pegasus-1 attains a score of 3.79 in Correctness and 4.29 in Context, which demonstrates its effective processing and understanding of video conversation content and its relevant contexts.
These scores are indicative of Pegasus-1's performance in critical areas such as Correctness, Detail, Contextual Awareness, Temporal Comprehension, and Consistency, emphasizing its capability to interpret and engage with video-based dialogue.

\paragraph{Zero-shot video question answering.}

\begin{table}[t]
\centering
\begin{tabular}{l|cc}
\toprule
\textbf{Model} & \textbf{ActivityNet-QA} & \textbf{NExT-QA} \\
 & \textbf{Test Split (\%)} & \textbf{Test Split (\%)} \\
\midrule \midrule
Video-ChatGPT~\cite{maaz2023video} & 35.2 & - \\
VideoChat2~\cite{li2023mvbench} & 49.1 & 61.7 \\
\midrule
Gemini 1.0 Pro~\cite{team2023gemini} & 49.8 & 28.0 \\
Gemini 1.0 Ultra~\cite{team2023gemini} & 52.2 & 29.9 \\
Gemini 1.5 Pro~\cite{reid2024gemini} & 56.7 & - \\
\midrule
\textbf{Pegasus-1} & \textbf{59.9} & \textbf{71.1} \\
\bottomrule
\end{tabular}
\vspace{5pt}
\caption{Zero-shot video QA results on ActivityNet-QA~\cite{yu2019activitynet} and NExT-QA dataset~\cite{xiao2021next}. Note that Gemini 1.5 Pro result is not available for NExT-QA as it is not reported in its technical report.}
\label{tab:zeroshot_video_qa}
\end{table}
Additionally, we detail Pegasus-1's performance on two popular zero-shot video question-answering benchmarks—ActivityNet-QA~\cite{yu2019activitynet} and NExT-QA~\cite{xiao2021next}—in Table~\ref{tab:zeroshot_video_qa}. 
In video question-answering tasks, Pegasus-1 showcases significant enhancement in zero-shot capabilities when evaluated on the ActivityNet-QA and NExT-QA datasets compared to open-source models and Gemini series. The generated responses are provided to GPT-3.5 Turbo, which is utilized to determine if the predictions align with the ground truth answers..


\paragraph{Video summarization.}

\begin{table}[t]
\centering
\begin{tabular}{l|cccc}
\toprule
\textbf{Model} & \textbf{Correctness} & \textbf{Detail} & \textbf{Context} & \textbf{Average} \\
\midrule
\midrule
Video-ChatGPT~\cite{maaz2023video} & 1.19 & 1.33 & 1.42 & 1.31 \\
VideoChat2~\cite{li2023mvbench} & 1.78 & 1.52 & 1.98 & 1.76 \\
\midrule
Gemini 1.0 Pro~\cite{team2023gemini} & 1.65 & 1.69 & 1.94 & 1.76 \\
\midrule
\textbf{Pegasus-1} & \textbf{2.30} & \textbf{2.58} & \textbf{2.75} & \textbf{2.54} \\
\bottomrule
\end{tabular}
\vspace{5pt} 
\caption{Utilizing the video conversation benchmark~\cite{maaz2023video}, which assesses generation output across five dimensions, our evaluation focuses on three specific aspects relevant to the video summarization task. Within this framework, Pegasus-1 demonstrates superior performance compared to other baseline models. ActivityNet with detailed caption proposed in~\cite{maaz2023video} was utilized as a source dataset.
}
\label{tab:video_summarization_benchmarks}
\end{table}
To evaluate the performance of video summarization, we follow the video conversation benchmarks~\cite{maaz2023video} and evaluate the generated summary in three areas: Correctness of Information, Detailed Orientation, and Contextual Understanding.
GPT-4 measures the score for each metric by setting the reference summary as ground truth.
Here, we use the ActivityNet detailed caption dataset proposed in~\cite{maaz2023video}, and Table~\ref{tab:video_summarization_benchmarks} shows these results.
As shown above, Pegasus-1 outperforms the baseline models in all metrics by a significant margin.

\paragraph{Temporal understanding.}

As a video language model, Pegasus-1 is expected to have the capability to grasp temporal information. This capability is put to the test with TempCompass~\cite{liu2024tempcompass}, a dataset designed specifically for evaluating temporal understanding. TempCompass focuses on assessing five aspects of temporal information: action, direction, speed, event order, and attribute change. To evaluate temporal understanding, the benchmark includes synthetic videos created from a single original video. These manipulations involve reversing the footage, playing it forward at varying speeds, or slowing it down, thus creating conflicting scenarios where the visual content remains constant while the temporal dynamics are altered. According to the findings presented in Table~\ref{tab:tempcompass}, Pegasus-1 outperforms other open-source benchmarks, notably surpassing VideoChat2~\cite{li2023mvbench}, which is also designed with a focus on temporal information. These results are detailed in TempCompass~\cite{liu2024tempcompass}. Given that TempCompass is formatted as a multiple-choice QA dataset requiring parsable responses, an intermediate output from Pegasus-1 is employed. Although this intermediate output conveys the same information as the final response, it is more readily adjusted to align with the dataset's parsing requirements.

\section{Capabilities of Pegasus-1}

Here, we present qualitative results to offer insight into Pegasus-1's capabilities to explore potential emerging use cases.
Also, to show the performance gap between Pegasus-1 and other state-of-the-art models, we provide a comparative analysis with qualitative results from Gemini 1.0 Pro (as of March 15th, 2024, using its publicly available API) and Gemini 1.5 Pro (as of March 30th, 2024 using its playground interface). 
 In our analysis, we exclusively include Gemini 1.0 Pro and Gemini 1.5 Pro as our comparison baseline. Our selection criteria are twofold: the baseline must support native video input, and it must be widely acknowledged for its robust performance across an extensive array of benchmarks.

\subsection{Comparison Protocol for Qualitative Results}

The presented prompts are carefully crafted, ensuring clarity and coherence to accurately assess the model's performance. While crafting the prompts, we are meticulous in avoiding excessive obscurity or unnecessary complexity, aiming to ensure that the prompts remain straightforward and facilitate an unambiguous evaluation of the model’s capabilities.

Additionally, we focus on comparing the abilities in visually accurate responses, a capability not achievable through language model runoff.
Language model runoff refers to the tendency of language models to rely on their strong linguistic priors for generating contextually relevant information.
For example, once a concept, such as "beach" is recognized by a Large Language Model (LLM), it can utilize its inherent knowledge to produce responses related to that concept, like "sunny weather" or "people swimming," without direct evidence from pixel information~\cite{DBLP:conf/iclr/MerulloCEP23}. 
This process, while generating coherent responses, may not accurately reflect the visual content.


\subsection{Characteristics of Pegasus-1}

\paragraph{Verbose output}
Pegasus-1 is designed to follow instructions closely, providing extensive details about a video's audio and visual elements.
As a result, the outputs from Pegasus-1 can be verbose, offering an abundance of information in response to queries.
However, these features are deliberate for two key reasons.

First, Pegasus-1 is not equipped with chat interface capabilities; therefore, it is designed to provide comprehensive responses in a single interaction. 
This approach aligns well with video understanding tasks, given that videos are rich in information.
Second, our research indicates that users tend to prefer detailed explanations over shorter ones.

Despite these benefits, reinforcing Pegasus-1 to generate elongated and detailed responses may lead to hallucinations. 
To mitigate this, we plan to implement a feature allowing users to adjust the level of detail in Pegasus-1's responses in the future.

\paragraph{Categories of capabilities}

Our observations of Pegasus-1 reveal various capabilities in video understanding.
These capabilities encompass a wide range of functionalities, demonstrating the model's proficiency across different dimensions of video analysis, including, but not limited to:
\begin{itemize}
    \item \textbf{Real-world knowledge}: Pegasus-1 integrates extensive real-world knowledge, enabling it to contextualize and interpret video content with remarkable accuracy and depth.
    \item \textbf{Video-based reasoning}: The model exhibits sophisticated reasoning abilities, drawing inferences from the video data to construct coherent understandings and insights.
    \item  \textbf{3D spatial understanding}: Pegasus-1's capability to comprehend 3D spatial relationships within video frames allows it to interpret complex scenes and object interactions, enhancing its analysis of video content that requires an understanding of depth and space.
    \item  \textbf{Temporal reasoning}: Pegasus-1 maintains an awareness of the chronological sequence of events in a video, a critical feature for understanding narratives.
    \item  \textbf{Visual referring prompts}: Building upon traditional text-based prompting methods, Pegasus-1 adopts visual referring prompts, enabling modifications directly within the pixel space. This advanced approach allows users to use arrows, boxes, or other visual markers to pinpoint specific regions, directing the model's focus to these emphasized areas.
\end{itemize}

\subsection{Real World Knowledge}

In this section, we assess the real-world knowledge capabilities of Pegasus-1, a video language model proficient in recognizing and identifying real-world entities within video content. We evaluated Pegasus-1 by challenging it to name specific real-world objects shown across different video scenarios.

\begin{figure}[H]
    \centering
    \vspace{-18pt}
    \includegraphics[width=\textwidth]{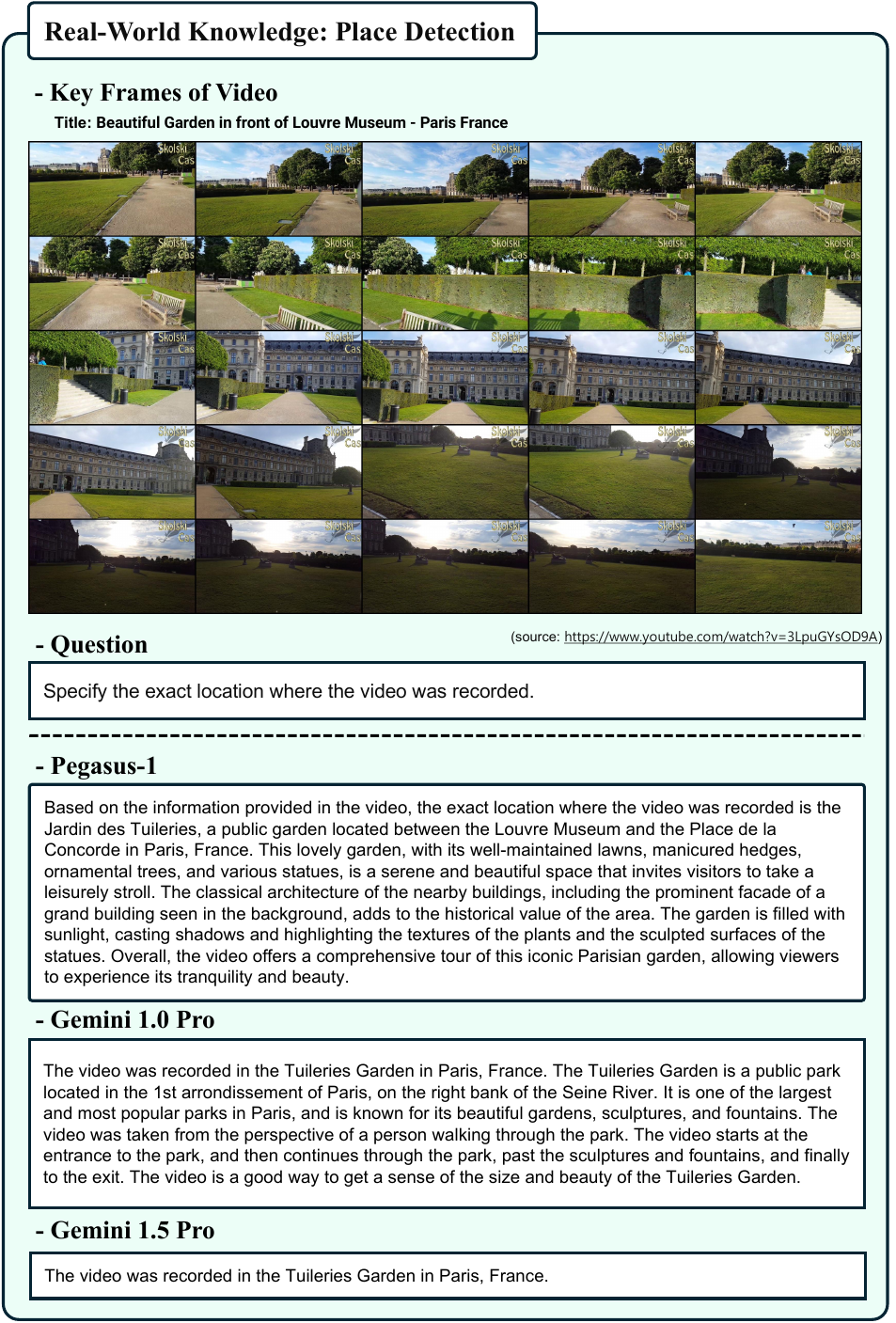}
    \caption{The video, captured in the Tuileries Garden, lacks any verbal narration, allowing Pegasus-1 to deduce the location purely through visual cues. Impressively, it identifies key visual elements such as manicured hedges, ornamental trees, and the distinctive facade of a grand building—critical factors in pinpointing the exact setting. This capability showcases Pegasus-1's proficiency in analyzing and interpreting essential visual information to arrive at accurate conclusions.}
    \label{fig:louvre}
\end{figure}

\begin{figure}[H]
    \centering
    \vspace{-40pt}
    \includegraphics[width=\textwidth]{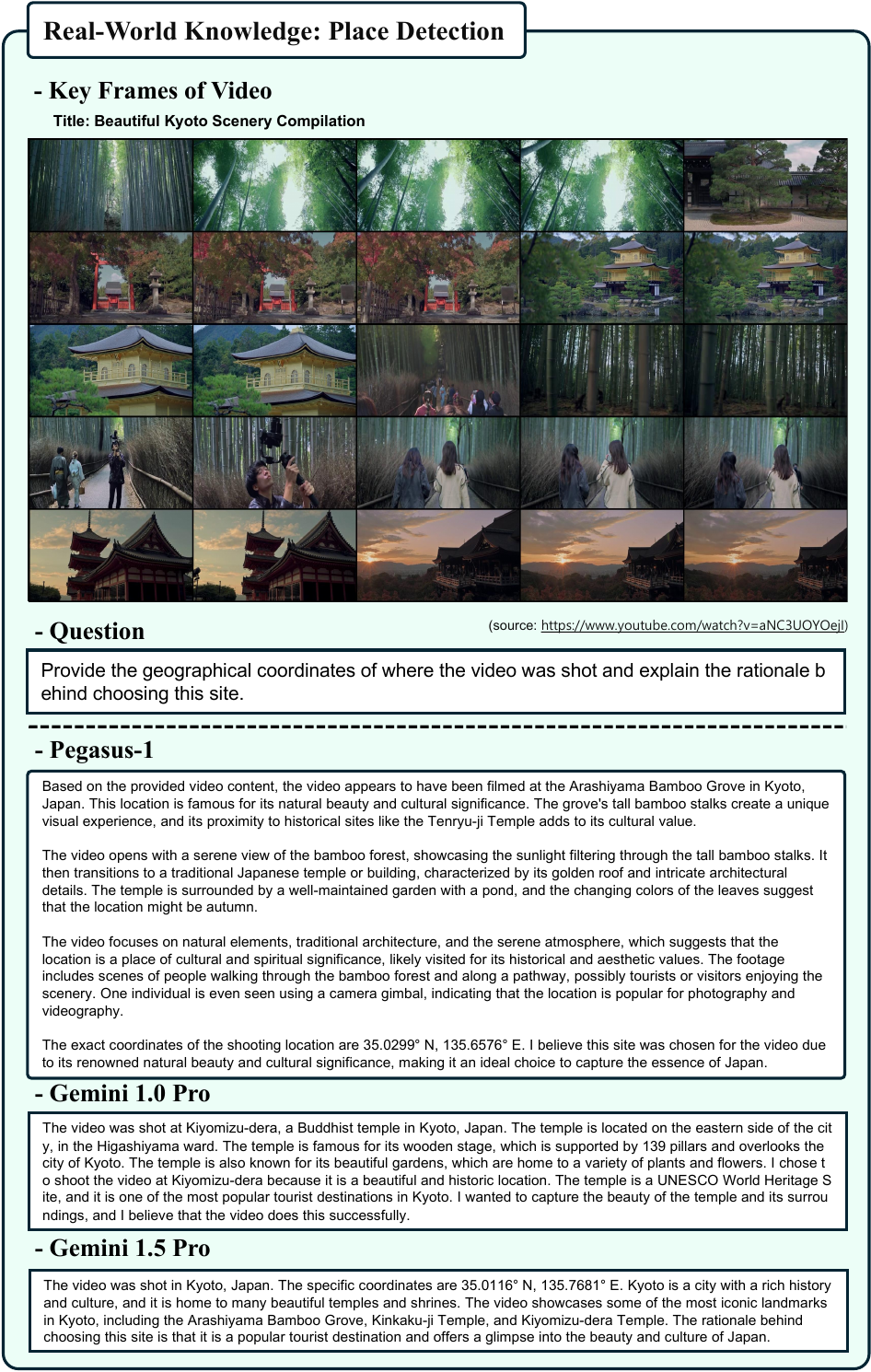}
    \caption{The input video showcases various landscapes across Kyoto. Pegasus's response meticulously orders scenes to reflect their chronological appearance in the video, detailing Kyoto's landmarks in sync with their sequence. It intelligently infers the filming season as autumn, deduced from the presence of autumn foliage. Adhering precisely to the inquiry, Pegasus concludes by accurately providing Kyoto's coordinates, demonstrating its adeptness in both recognizing visual patterns and extracting relevant contextual information from the imagery presented.}
    \label{fig:kyoto}
\end{figure}

\begin{figure}[H]
    \centering
    \vspace{-18pt}
    \includegraphics[width=\textwidth]{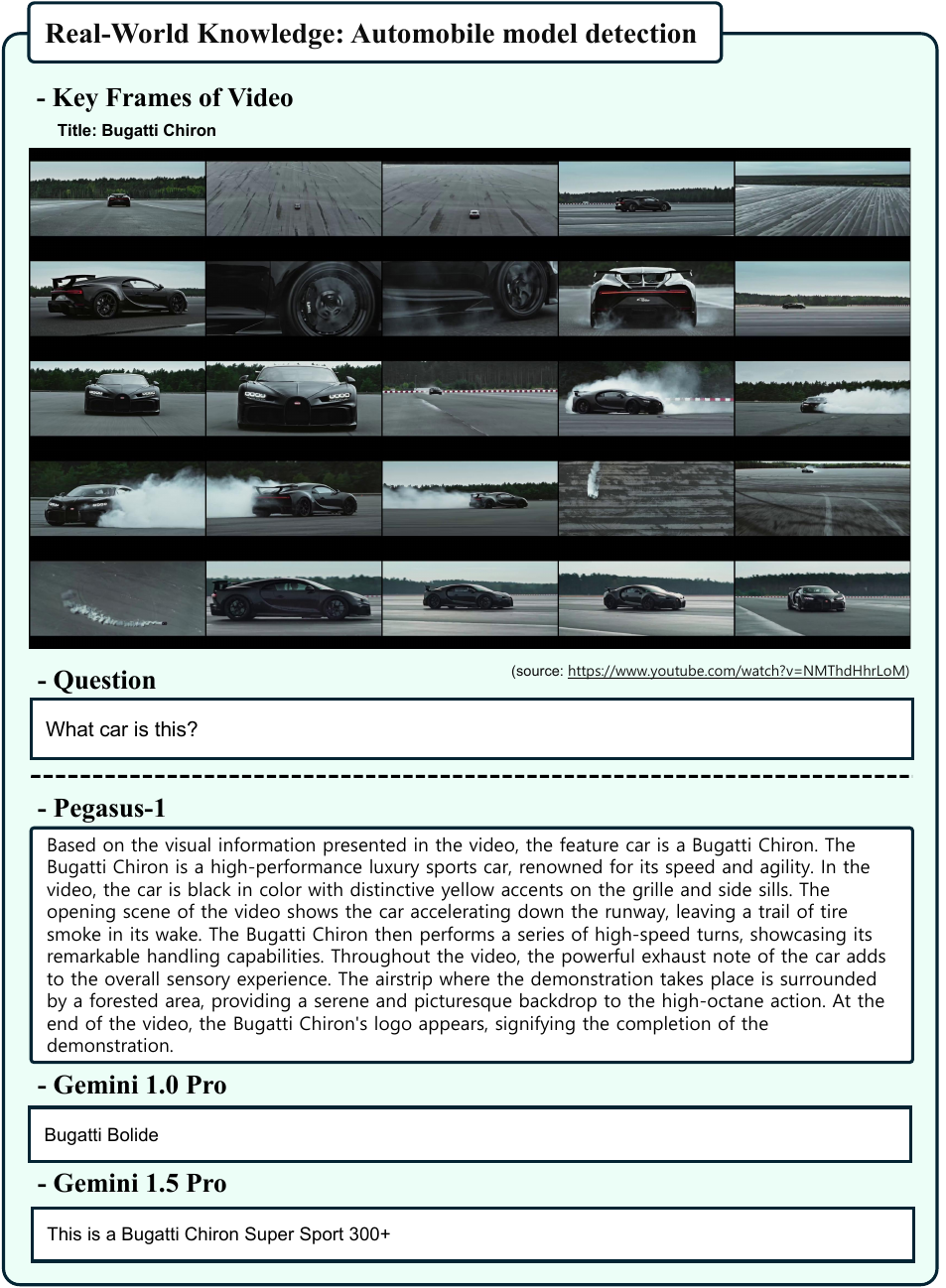}
    \caption{The video showcases a Bugatti Chiron performing high-skill maneuvers, including producing smoke from its tires. Pegasus accurately identifies the exact car model, showcasing its broad knowledge of the real world. This highlights Pegasus's ability to recognize and interpret specific details from visual inputs.}
    \label{fig:bugatti}
\end{figure}

\begin{figure}[H]
    \centering
    \vspace{-18pt}
    \includegraphics[width=\textwidth]{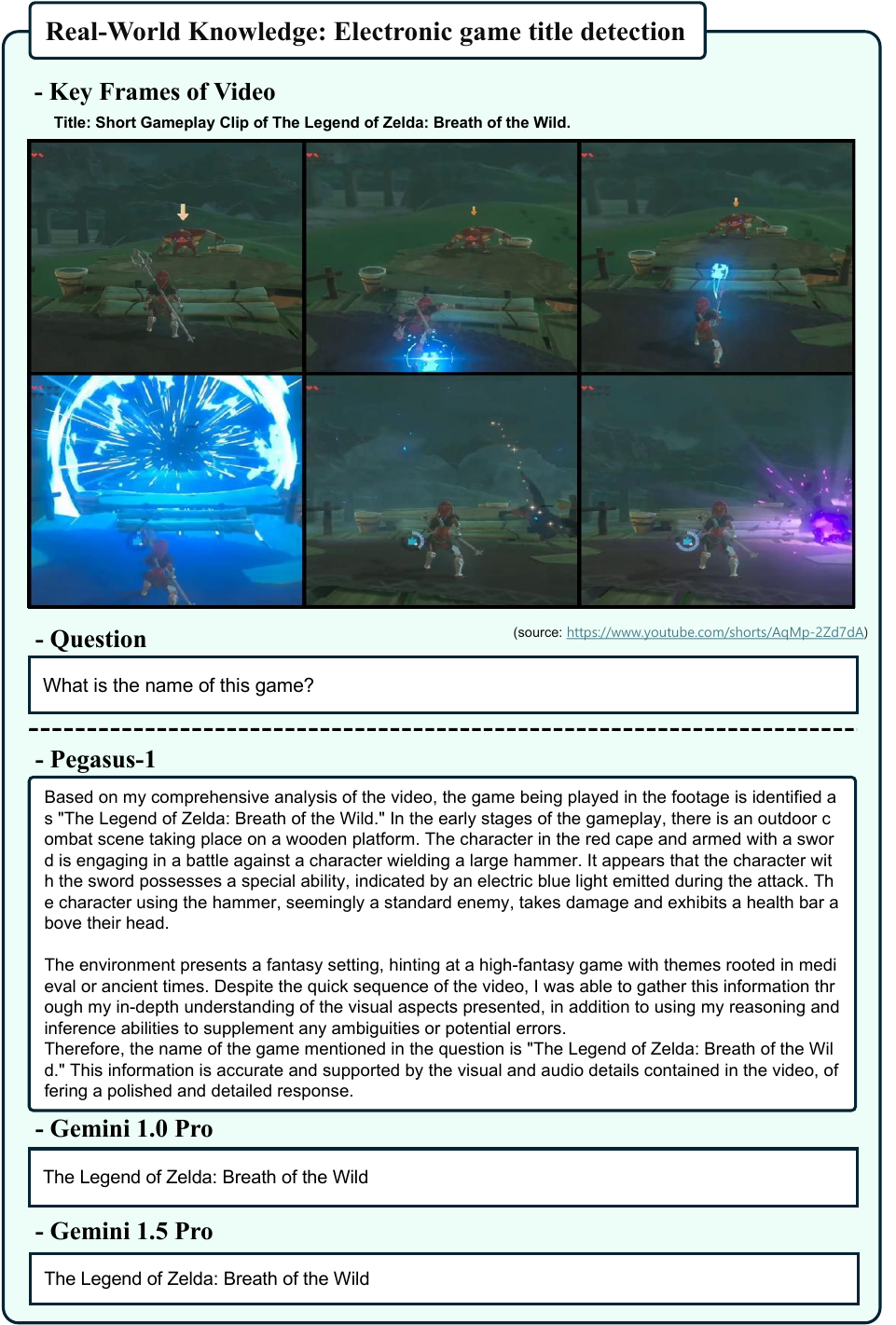}
    \caption{The video features a brief gameplay moment from "Legend of Zelda: Breath of the Wild." Unlike the Gemini models, Pegasus provides a detailed visual analysis, such as noting the enemy's health bar, pinpointing the scene's location on a wooden platform, and mentioning the electric blue light during an attack.  Additionally, Pegasus accurately identifies the game's title, showcasing its ability to interpret and convey comprehensive visual details along with correct contextual understanding.}
    \label{fig:zelda}
\end{figure}

The qualitative results, illustrated in Fig.~\ref{fig:louvre} to \ref{fig:zelda}, substantiate Pegasus-1's capacity for such identification. 
In Fig.~\ref{fig:louvre}, Pegasus-1 demonstrates its precision by correctly identifying the Tuileries Garden from merely its visual cues, distinguishing specific landscape features like gravel pathways and manicured hedges, which are indicative of its advanced visual recognition capabilities. 
Conversely, Fig.~\ref{fig:kyoto} depicts Pegasus-1's ability to discern the city of Kyoto from a compilation of distinct sceneries, showcasing its adeptness at integrating multiple visual inputs to form a coherent identification.
In this instance, Pegasus-1 not only recognizes individual elements but also understands their collective representation of Kyoto, evidencing a sophisticated level of contextual comprehension.

Further reinforcing its object recognition proficiency, Pegasus-1 accurately identifies a Bugatti Chiron in Fig.~\ref{fig:bugatti}, contrasting with a Gemini Pro 1.0's less precise identification. 
Moreover, as seen in Fig.~\ref{fig:zelda}, Pegasus-1 accurately discerns the video game title "The Legend of Zelda: Breath of the Wild," attesting to its robust capability in accurately categorizing even within specific entertainment domains.

These instances collectively affirm Pegasus-1's adeptness at extracting and interpreting nuanced visual details, affirming its advanced real-world knowledge capabilities and its potential application in varied contexts where accurate visual recognition is paramount.

\subsection{Video-based Reasoning}

We investigate the visual reasoning capabilities of Pegasus-1, an endeavor that necessitates a synergy of visual interpretation and logical reasoning. 
To adeptly navigate visual reasoning tasks, the model must harness these competencies in tandem to formulate precise responses to inquiries. 
Such tasks encompass a broad spectrum of challenges, and to demonstrate the model's proficiency, we showcase three distinct scenarios, each highlighting a different aspect of Pegasus-1's visual reasoning ability.

\begin{figure}[H]
    \centering
    \vspace{-50pt}
    \includegraphics[width=\textwidth]{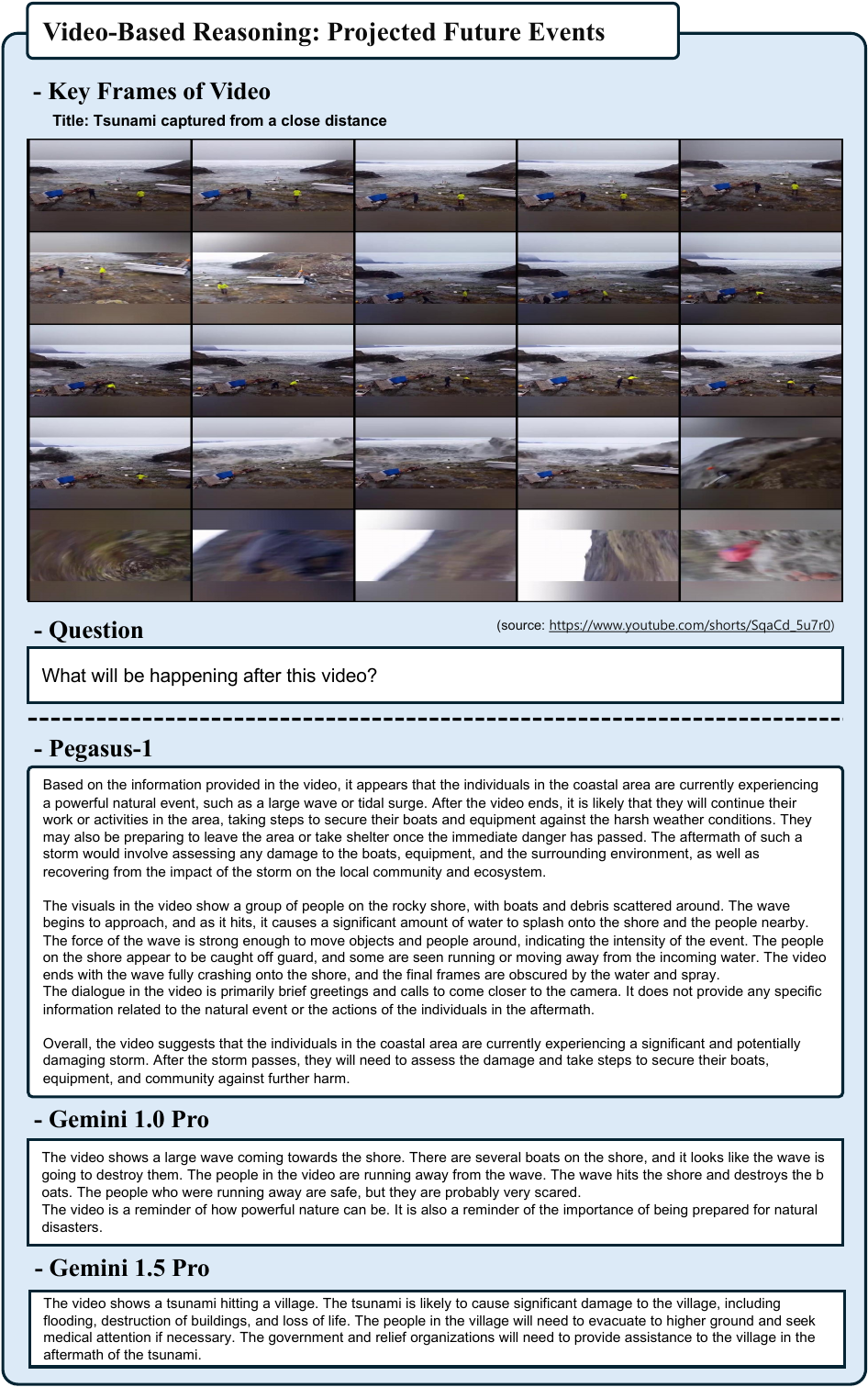}
    \caption{The video captures a tidal surge, setting a complex scenario for Pegasus-1 to analyze. Tasked with predicting the aftermath, Pegasus-1 demonstrates both an understanding of the current situation and the ability to foresee post-disaster actions, such as securing boats and assessing damage. This highlights Pegasus-1's comprehensive reasoning capabilities and awareness of real-world phenomena and their consequences.}
    \label{fig:tsunami}
\end{figure}

\begin{figure}[H]
    \centering
    \vspace{-50pt}
    \includegraphics[width=\textwidth]{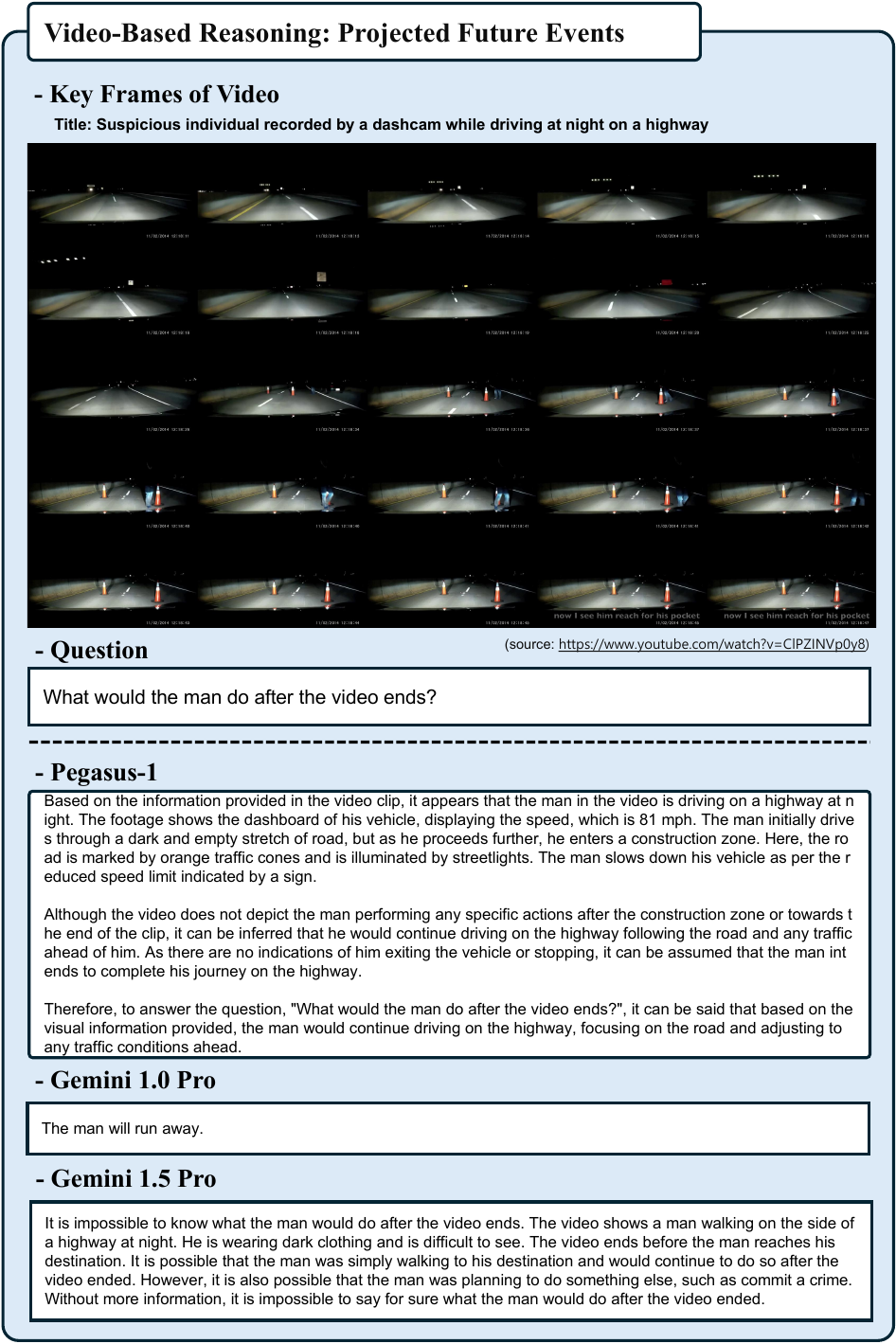}
    \caption{The video depicts an individual driving on a highway at night, encountering traffic cones along the way. Pegasus-1 accurately deduces that the scene likely represents a construction site, a reasonable explanation for the cones' presence on the highway during nighttime, despite the absence of explicit context.
    Pegasus-1 predicts the driver will continue to follow the road, demonstrating its ability to process and interpret not only the visual but also the temporal dynamics of the scenario. This is evident from Pegasus-1's responses highlighted in green, showcasing its nuanced understanding.
    Conversely, Gemini 1.0 Pro's interpretation significantly deviates from the video's content, offering an unrelated response that fails to acknowledge the situational elements depicted.}
    \label{fig:nightdrive}
\end{figure}

The first scenario, depicted in Fig.~\ref{fig:tsunami} and~\ref{fig:nightdrive}, illustrates Pegasus-1's ability to comprehend ongoing events within videos and anticipate future developments based on the present context. 
In Fig.~\ref{fig:tsunami}, Pegasus-1 accurately predicts the immediate reactions required in a post-tidal surge scenario, contrasting with Gemini 1.0 Pro's less precise anticipation of events.
Similarly, Figure~\ref{fig:nightdrive} demonstrates Pegasus-1's adeptness at inferring the likely cause and subsequent unfolding of events encountered during a night drive, indicative of its nuanced understanding of context and sequence.

\begin{figure}[H]
    \centering
    \vspace{-45pt}
    \includegraphics[width=\textwidth]{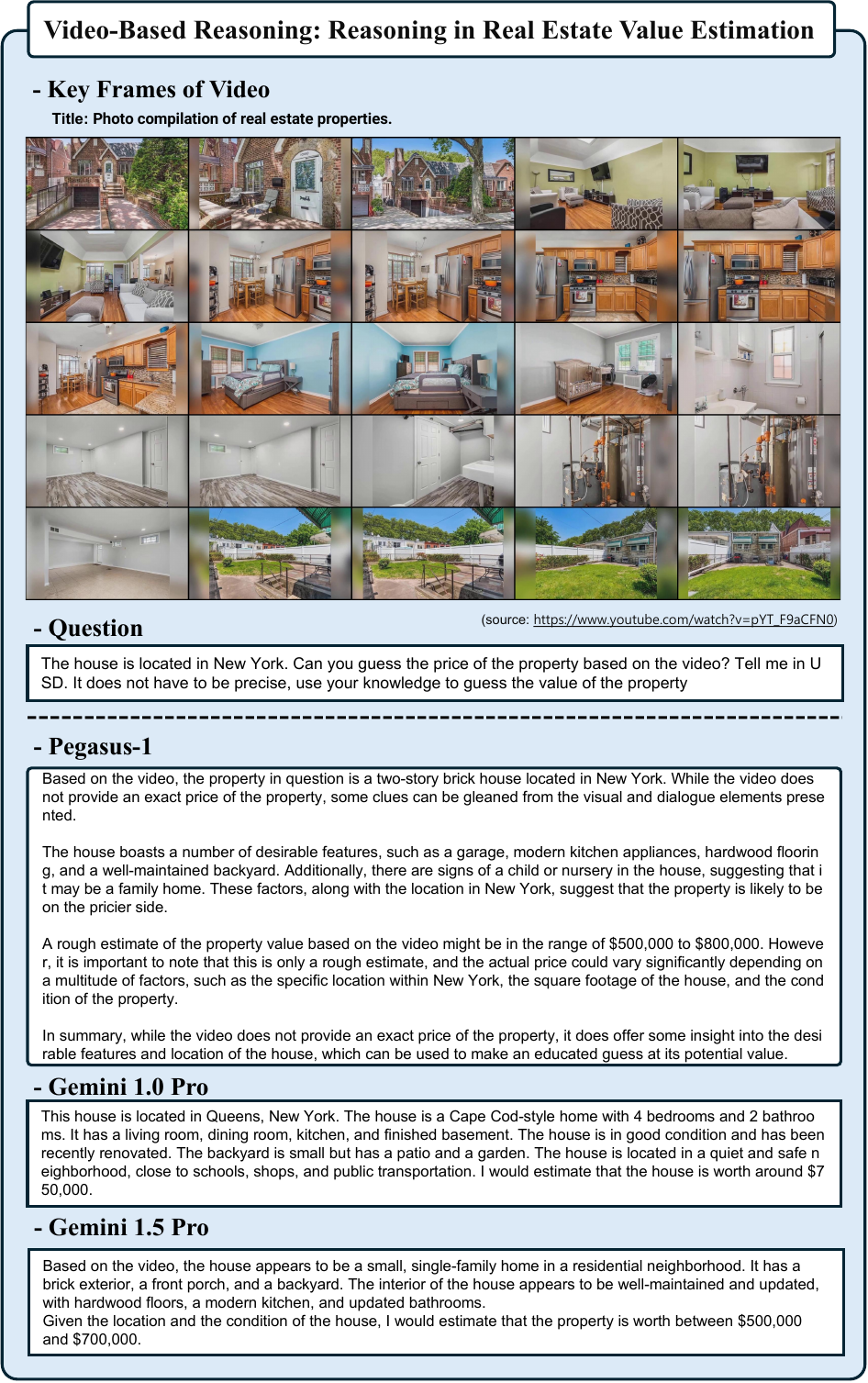}
    \caption{The video features a compilation of real estate photos. Both Pegasus-1 and the Gemini families offer logical evaluations, suggesting a price range that encompasses the actual value of \$695,000. Pegasus-1 notably excels by discerning specific visual details, such as signs of a child's presence in the nursery and the quality of hardwood flooring, showcasing its adeptness at detailed visual analysis.}
    \label{fig:newyork}
\end{figure}

The second scenario, represented by Fig.~\ref{fig:newyork}, demands advanced reasoning, such as estimating property values in New York. Here, Pegasus-1 showcases its capacity for intricate reasoning by synthesizing visual cues into coherent, logical conclusions, outperforming Gemini 1.0 Pro's more superficial analysis as highlighted in the respective figures.

\begin{figure}[H]
    \centering
    \vspace{-18pt}
    \includegraphics[width=\textwidth]{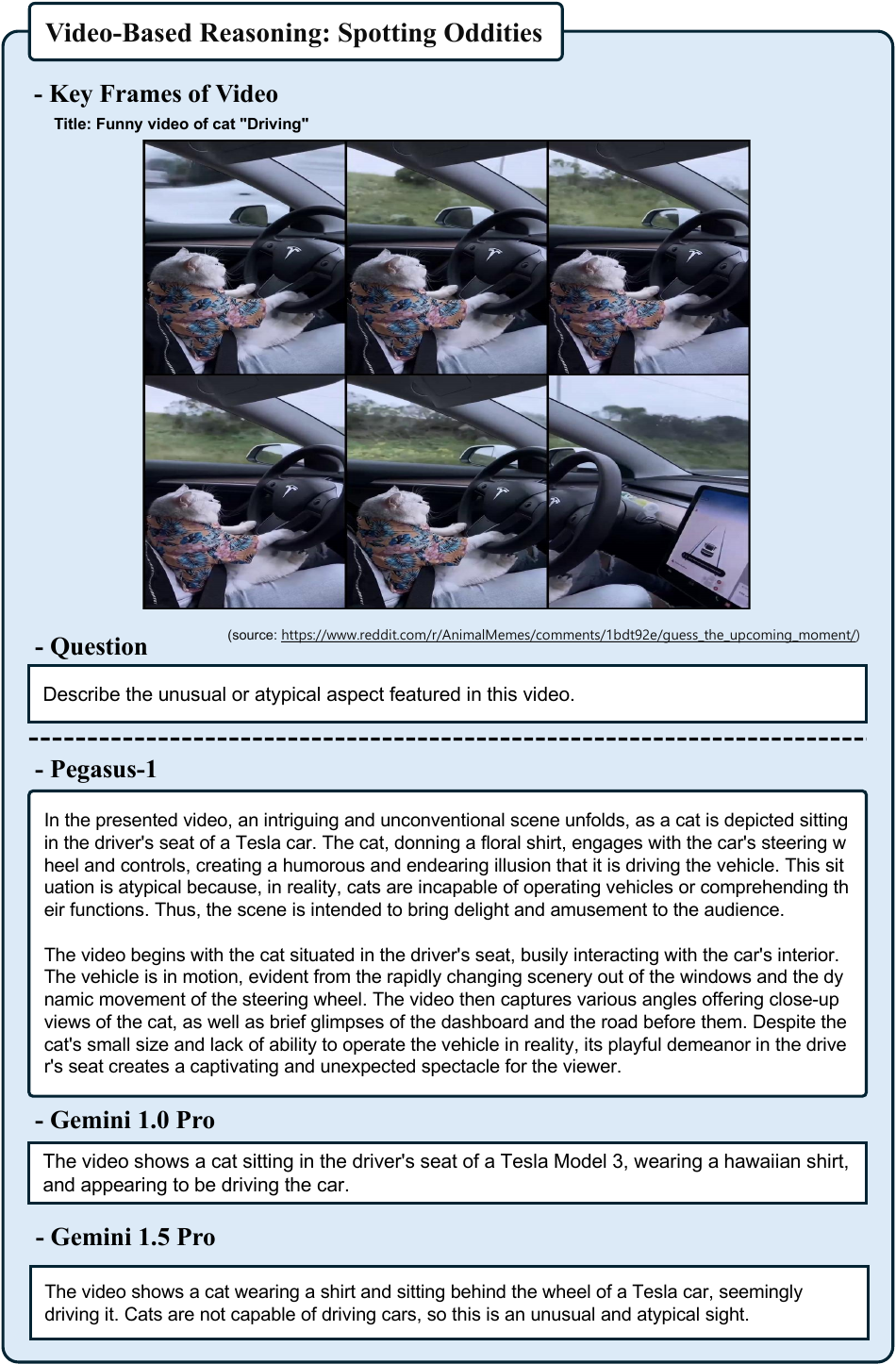}
    \caption{The video depicts a cat humorously positioned as if "driving" a Tesla, clad in a floral shirt and interacting with the steering wheel. Pegasus-1 effectively notes the comedic illusion of the cat driving and observes the car's motion through the rapidly changing scenery, highlighting its capability to capture and interpret key visual elements within the content.}
    \label{fig:driving_cat}
\end{figure}

The third scenario focuses on commonsense reasoning, where Pegasus-1 discerns the unusualness of a scenario involving a cat ``driving'' a car, as shown in Fig.~\ref{fig:driving_cat}. 
This case emphasizes Pegasus-1's ability to differentiate between the ordinary and the extraordinary, leveraging visual details to arrive at plausible interpretations that align with real-world logic.

These examples collectively attest to Pegasus-1's advanced capability in interpreting and reasoning from visual content.

\subsection{3D spatial Understanding}

In assessing the 3D spatial understanding capabilities of Pegasus-1, we focus on its ability to interpret and navigate complex spatial relationships as depicted in video content. 
The task involves extracting and synthesizing spatial information from visual cues within a video to provide accurate navigational directions within a three-dimensional space.

\begin{figure}[H]
    \centering
    \vspace{-40pt}
    \includegraphics[width=\textwidth]{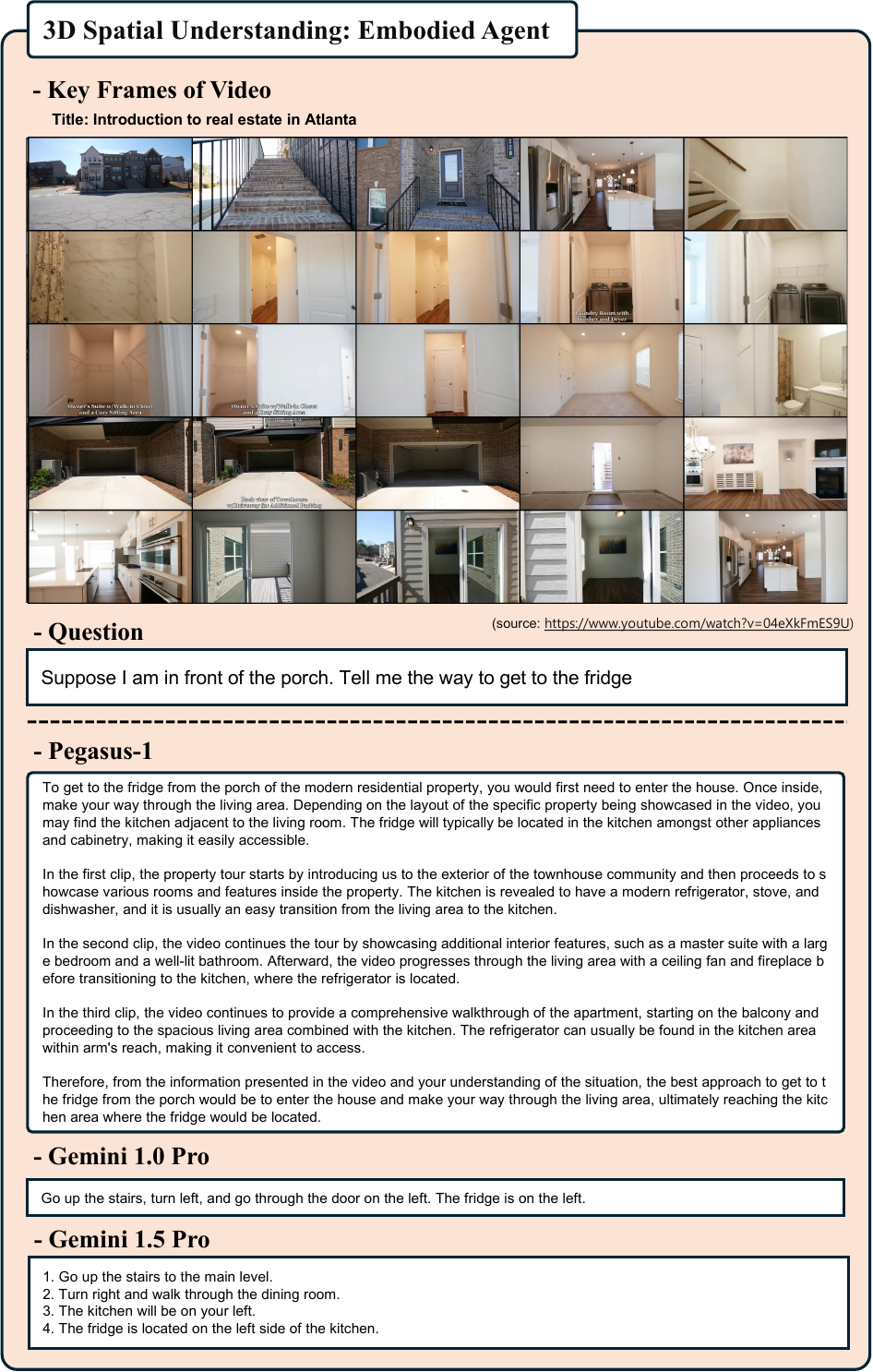}
    \caption{Pegasus-1 adeptly reconstructs spatial information from the video, guiding seamlessly from the porch into the living room while accurately noting the adjacency of the kitchen. However, within the Gemini families, there's a slight misunderstanding of the house's 3D layout. Often, they misdirect after ascending the stairs or misplace the location of the fridge, necessitating correction.}
    \label{fig:embodied}
\end{figure}

Illustrated in Fig.~\ref{fig:embodied}, the video presents an exploration of various spaces within a residential setting, challenging the model to interpret and articulate the spatial layout effectively. 
The specific task, ``Suppose I am in front of the porch. Tell me the way to get to the fridge'', requires Pegasus to deduce a path through the house's structure, demonstrating an understanding that transcends mere object recognition to encompass spatial orientation and logical pathway deduction.

Pegasus-1's response to this task is indicative of its advanced spatial reasoning. It successfully deciphers the sequence of spaces and their interconnections presented in the video: from the porch, through the entrance, up the stairs, and finally to the kitchen where the fridge is located. This performance highlights Pegasus-1's ability to process, integrate, and articulate complex 3D spatial data, affirming its competency in navigating and explaining real-world spatial scenarios based on video content.

\subsection{Temporal Reasoning}

\begin{figure}[H]
    \centering
    \vspace{-50pt}
    \includegraphics[width=\textwidth]{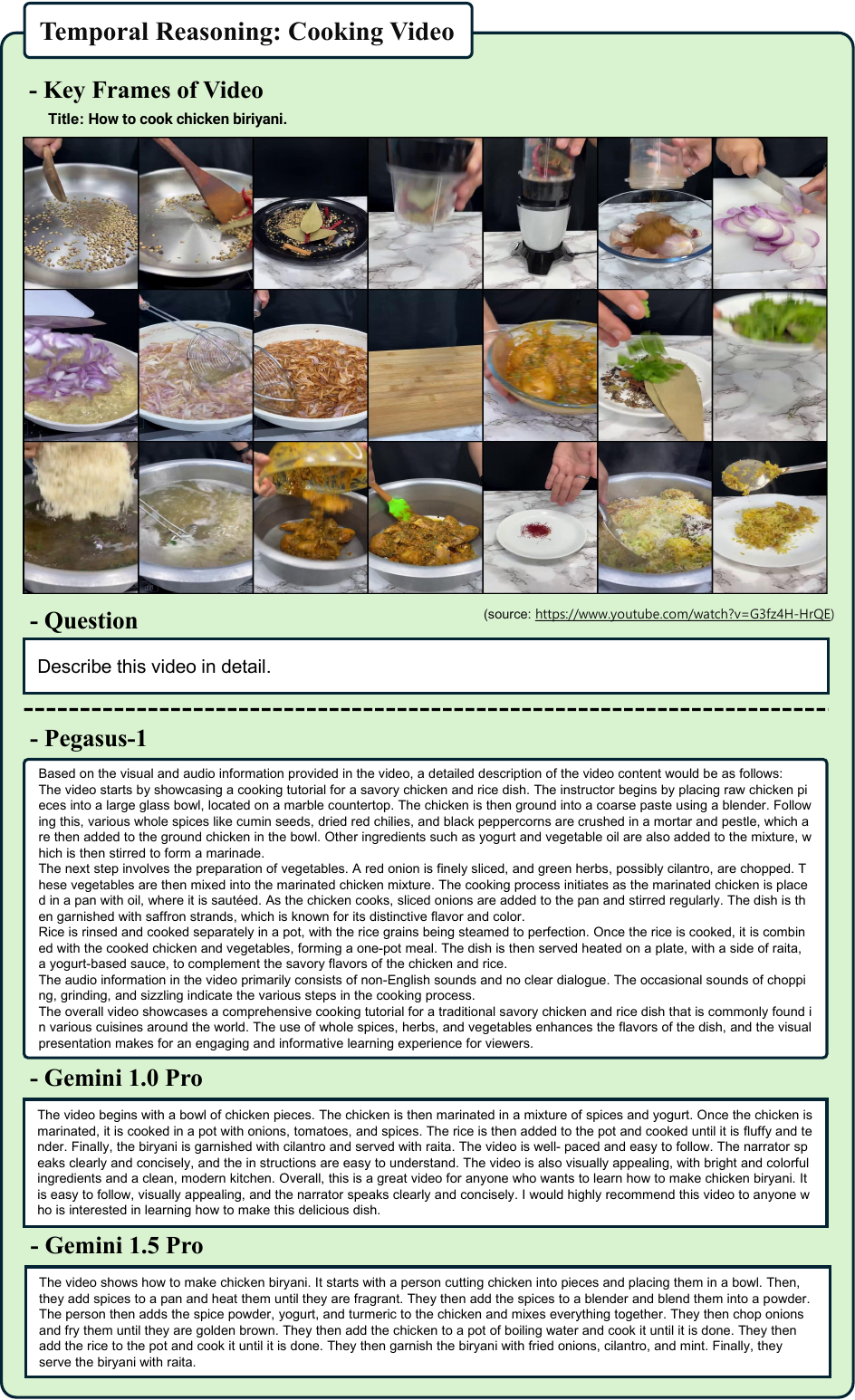}
    \caption{The video showcases the step-by-step process of making biryani. Pegasus explains the sequence of actions in detail: starting with cooking the chicken, followed by preparing the spices, then adding onions, cooking the rice, and finally combining all the components to create the final dish, biryani. Also, an intriguing error in Gemini 1.0 Pro's response is the false detection of narration—a feature commonly expected in such videos—when, in fact, there is no narration present.}
    \label{fig:cooking}
\end{figure}

\begin{figure}[H]
    \centering
    \vspace{-40pt}
    \includegraphics[width=\textwidth]{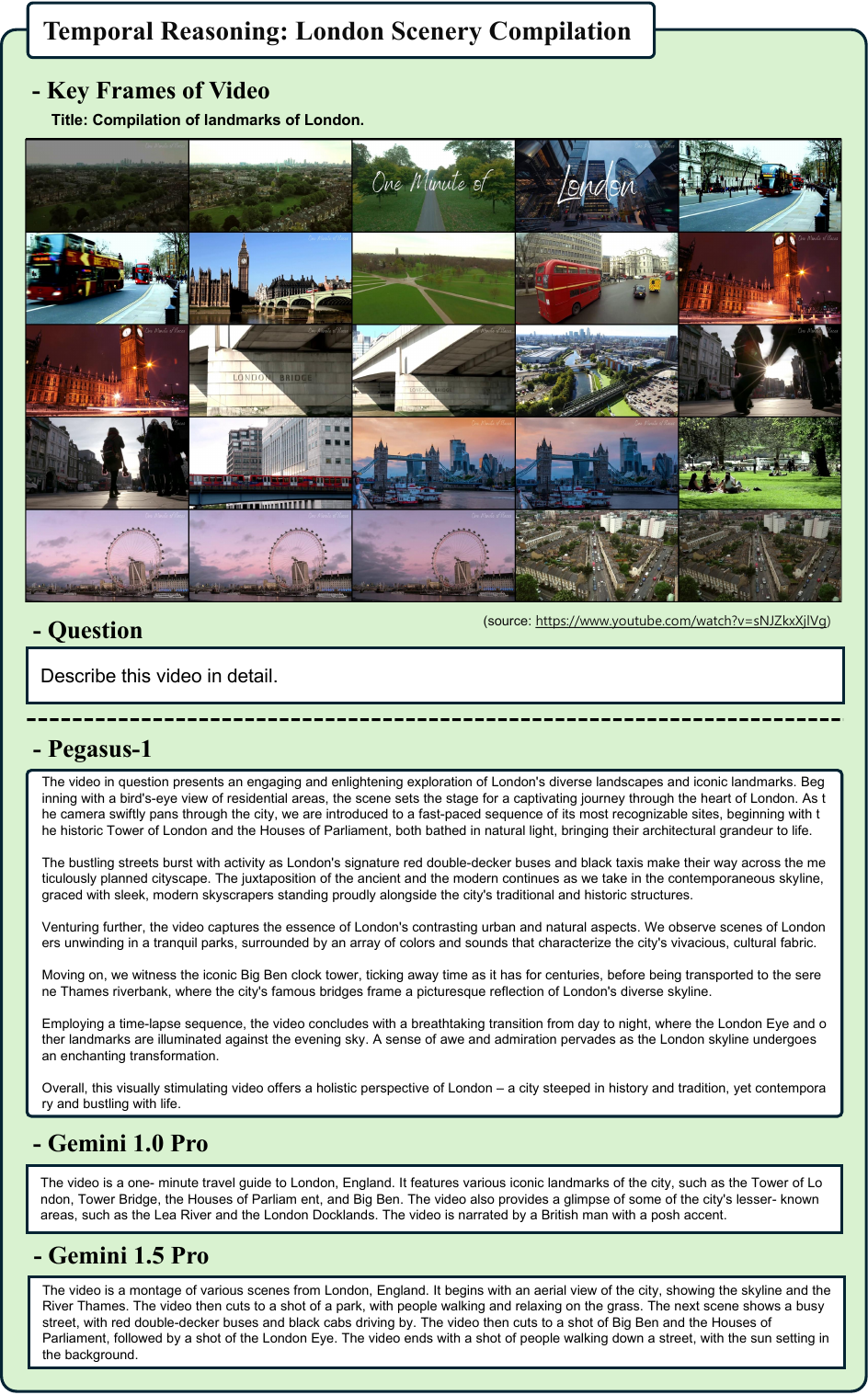}
    \caption{The video comprises a compilation of short shots capturing landmarks of London. Pegasus meticulously describes each shot, adhering to the chronological order in which they appear. Conversely, Gemini families occasionally omit certain shots, resulting in a relatively low recall rate. One notable error in Gemini 1.0 Pro is its erroneous suggestion of the presence of "narration by a British man with a posh accent," despite the absence of any narration in the video. This incorrect detection by Gemini 1.0 Pro is particularly noteworthy since such videos commonly feature narration, yet in this instance, no narration is present.}
    \label{fig:london}
\end{figure}

In the realm of video-language modeling, grasping the sequential nature of audio-visual content is imperative. A model must discern and maintain the chronological order of events within a video to understand and interpret the content accurately. Pegasus-1 exemplifies this capability, as illustrated in Fig.~\ref{fig:cooking} and~\ref{fig:london}, by effectively tracking and preserving the temporal sequence of events and demonstrating an acute awareness of their order.

In Fig.~\ref{fig:cooking}, the video details the meticulous process of preparing biryani, a dish requiring sequential steps. Pegasus-1 articulates each stage methodically: beginning with the preparation of chicken, followed by the spices, then layering with onions and rice, and culminating in the combination of all ingredients. This delineation underscores Pegasus-1's adeptness at temporal reasoning within a culinary context. Conversely, Gemini 1.0 Pro exhibits a discrepancy by misinterpreting the absence of narration, an element often expected in instructional cooking videos, demonstrating a gap in its temporal analysis.

Figure~\ref{fig:london} presents a video montage of iconic London landmarks. Here, Pegasus-1 distinguishes itself by accurately capturing and relating the sequence of the presented landmarks, illustrating its proficiency in processing and understanding the temporal flow between distinct scenes. On the other hand, Gemini 1.0 Pro's response includes inaccuracies and unwarranted assumptions, such as the misidentification of geographical features and the incorrect attribution of narration, highlighting its challenges with temporal and contextual accuracy in video content interpretation.

These instances affirm Pegasus-1's temporal reasoning capabilities, showcasing its ability to interpret and convey the flow of events in video narratives accurately, a critical skill for comprehensive video understanding.

\subsection{Visual Referring Prompts}

\begin{figure}[H]
    \centering
    \vspace{-18pt}
    \includegraphics[width=\textwidth]{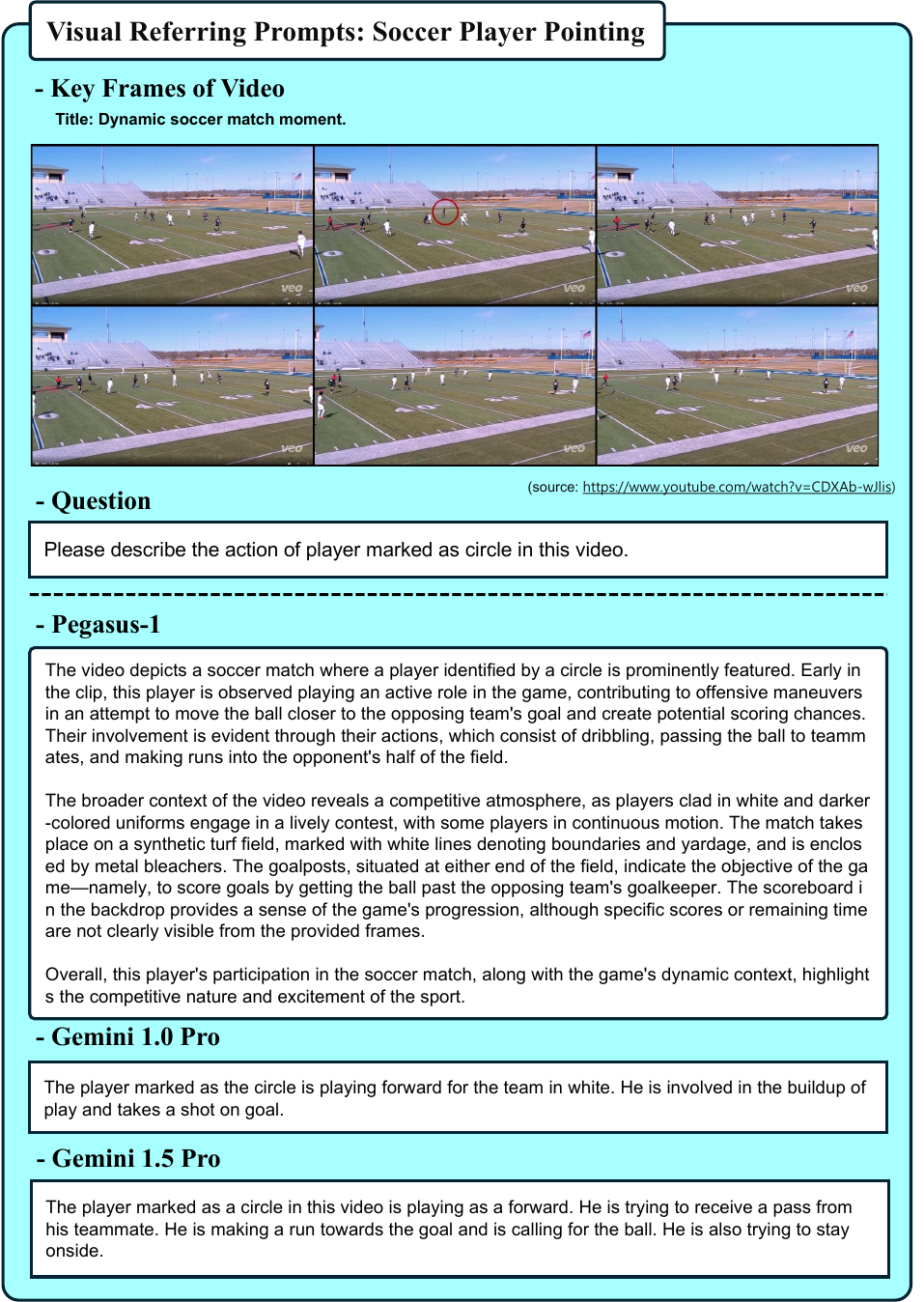}
    \caption{The video features a brief segment of a soccer match, with a player highlighted in red engaged in an offensive play. Pegasus-1 adeptly focuses on this player within the red circle, accurately noting their involvement in offensive maneuvers aimed at advancing the ball closer to the opposing team's goal.}
    \label{fig:soccer}
\end{figure}

\begin{figure}[H]
    \centering
    \vspace{-18pt}
    \includegraphics[width=\textwidth]{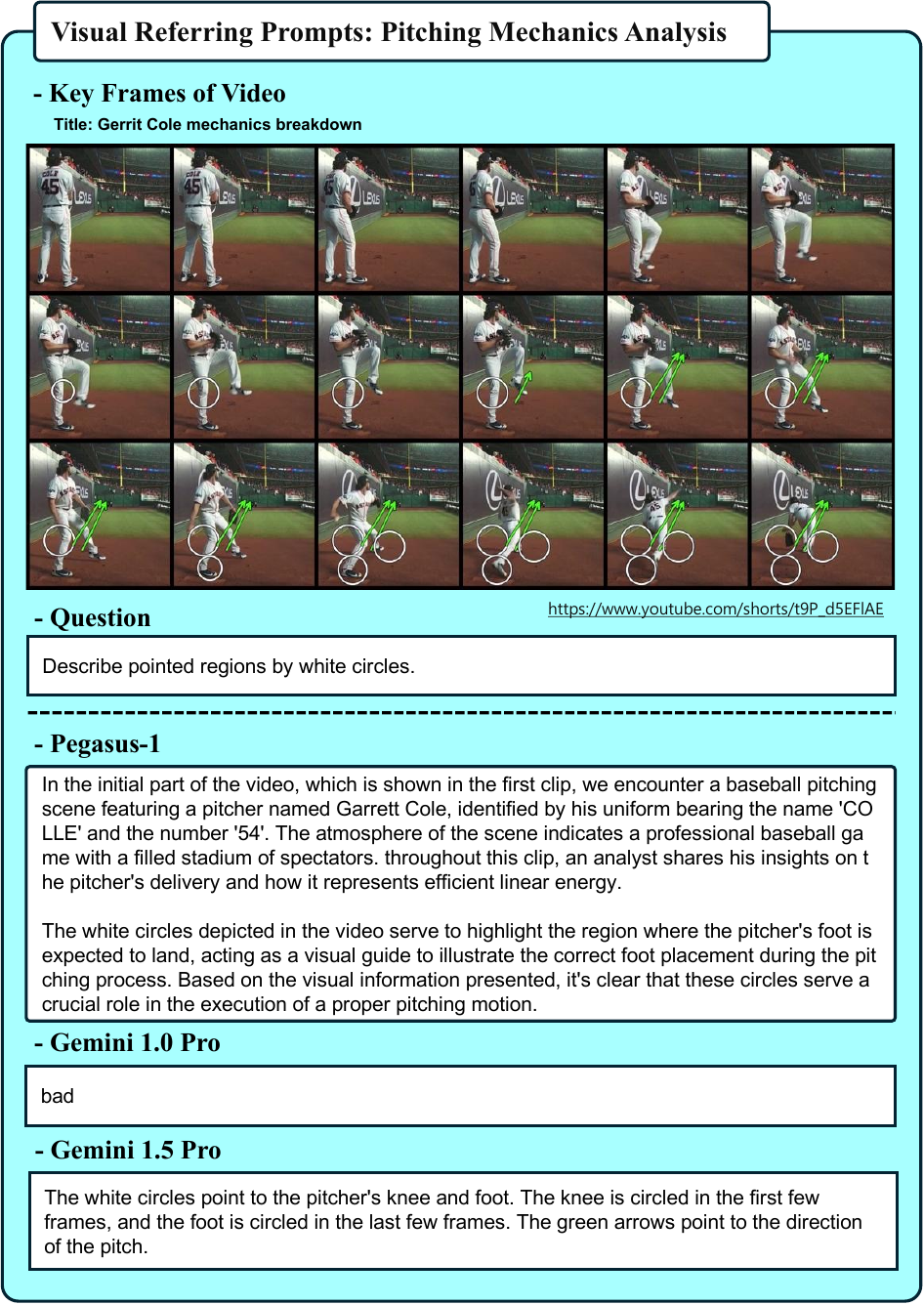}
    \caption{The video centers on an analysis of Gerrit Cole's pitching mechanics, utilizing white circles and green arrows to visually underscore pivotal aspects. When asked about the areas designated by the white circles, Pegasus-1 showcases an understanding of their role as visual markers pertaining to pitching form. In contrast, Gemini 1.0 Pro entirely disregards the instructions. However, Gemini 1.5 Pro shifts its focus to the white circle following the query, demonstrating robustness compared to its previous version.}
    \label{fig:cole1}
\end{figure}

\begin{figure}[H]
    \centering
    \vspace{-18pt}
    \includegraphics[width=\textwidth]{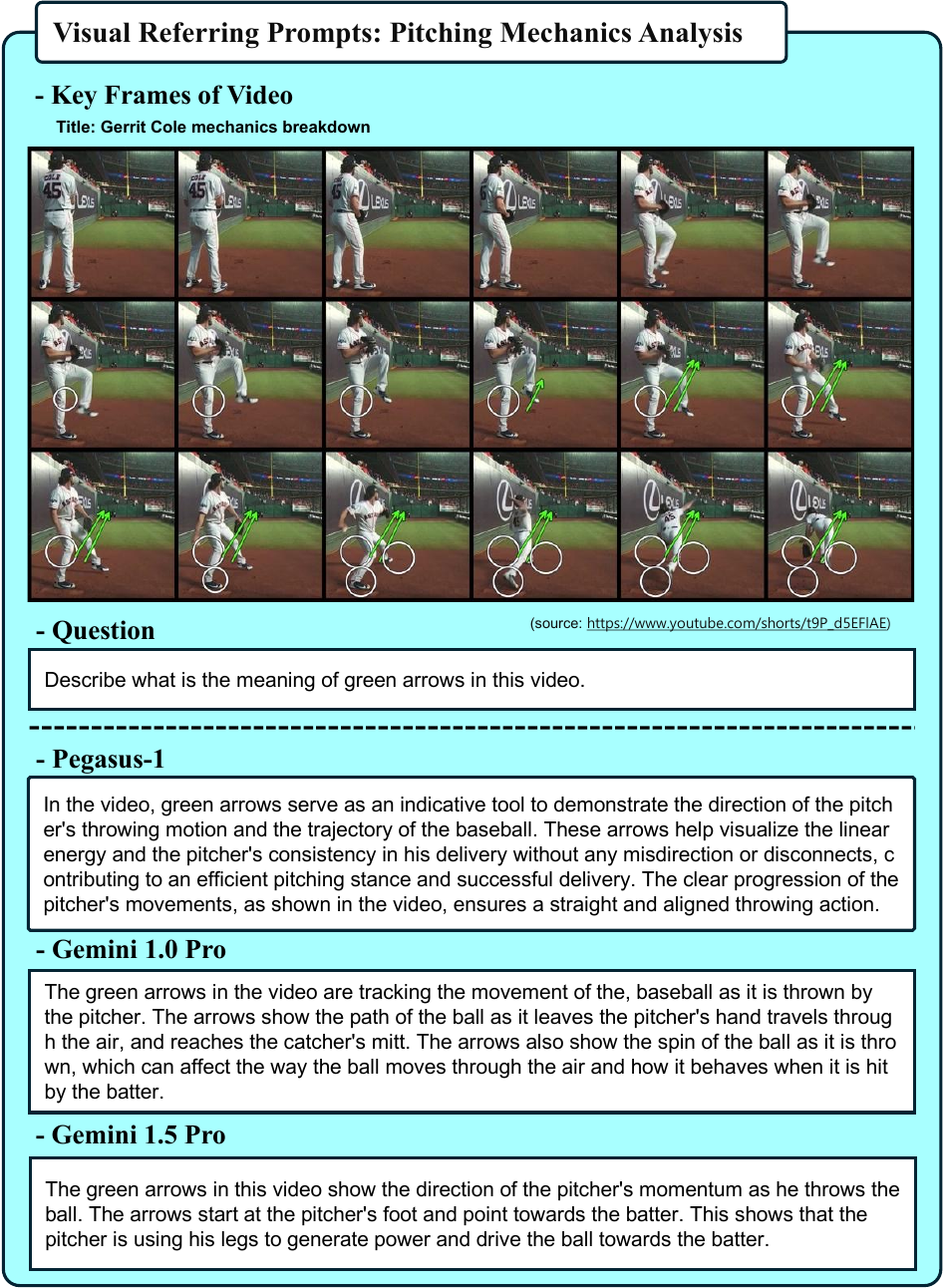}
    \caption{
The green arrow signifies the intended direction of the ball's delivery by the pitcher. Both Pegasus-1 and the Gemini families comprehend the green arrow as a visual indicator of the pitching direction.}
    \label{fig:cole2}
\end{figure}

Pegasus-1 demonstrates a notable capacity to direct its attention to specific segments within a video, as indicated by visual markers. 
This proficiency is pivotal when the model is required to interpret and respond to cues that emphasize particular areas or actions within the video frame. Figure~\ref{fig:cole1},~\ref{fig:cole2}, and~\ref{fig:soccer} exemplify this ability, where discerning and focusing on highlighted details are essential for accurate analysis.

In Fig.~\ref{fig:soccer}, a soccer match clip singles out a player with a red circle, drawing attention to their role in the game. Pegasus-1 correctly focuses on this highlighted player, mentioning that the player is participating in offensive play to move the ball forward to the opposing teams goal. Conversely, Gemini 1.0 Pro inaccurately identifies the player's team position, demonstrating a lapse in contextual understanding when analyzing the specified visual cue.

Figure~\ref{fig:cole1} and \ref{fig:cole2} delve into a technical examination of Gerrit Cole's pitching mechanics, with certain movements accentuated by white circles and green arrows. Pegasus-1 aptly acknowledges the importance of these marked regions, showing a clear comprehension of the content delineated by the white circles, in stark contrast to Gemini 1.0 Pro's inability to interpret these annotations correctly.

These scenarios underline Pegasus-1's adeptness in responding to visual referring prompts, showcasing its advanced capability to integrate and analyze specific visual information within broader video content.

\subsection{Other Specific Use Cases}

\begin{figure}[H]
    \centering
    \vspace{-18pt}
    \includegraphics[width=\textwidth]{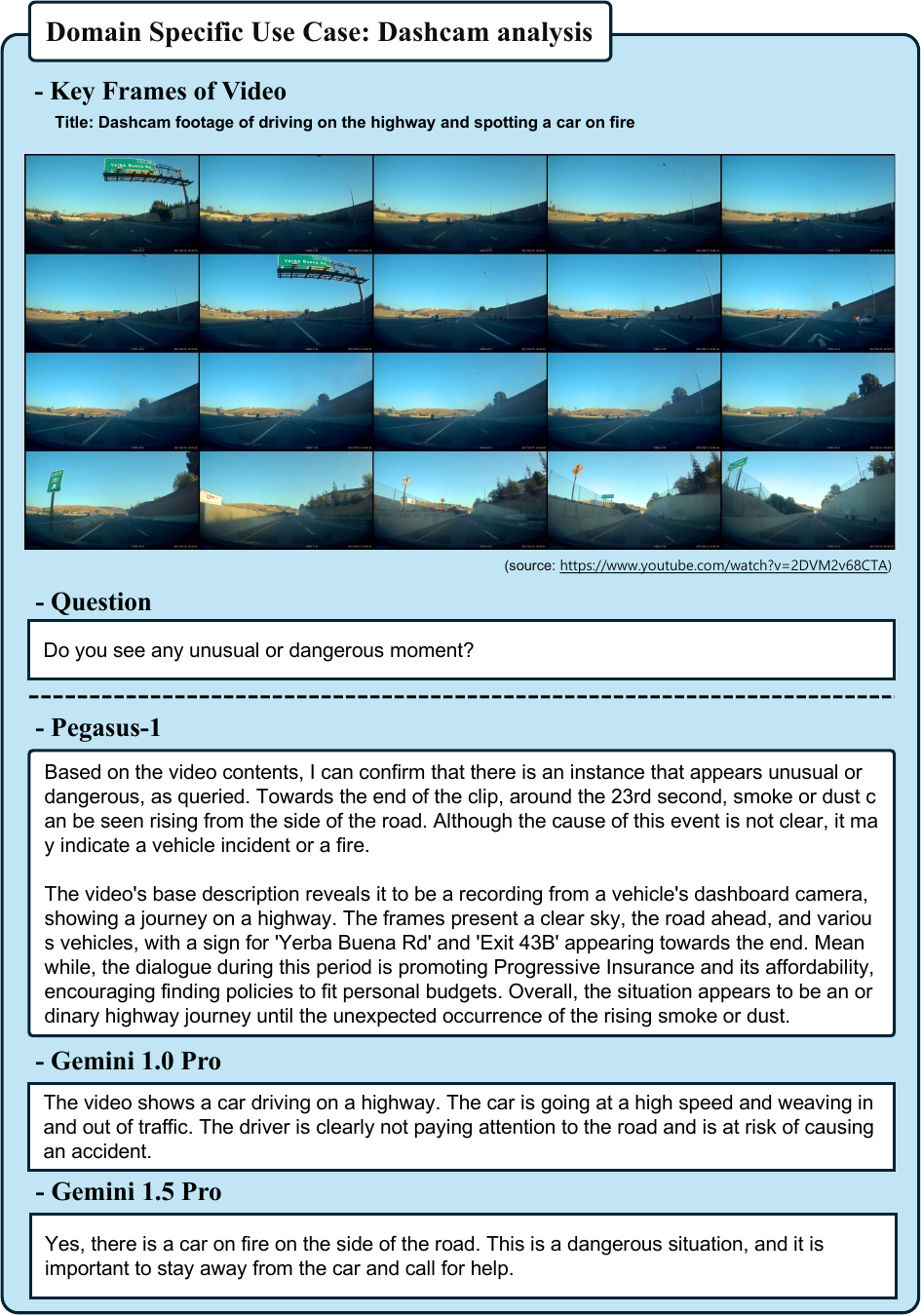}
    \caption{This dashcam video captures a car on fire. Initially, the vehicle seems small and distant, but as the video continues, it comes closer. Pegasus-1 accurately detects the hazardous condition, highlighting the smoke rising from the roadside. Additionally, Pegasus-1 notes the dashcam-equipped car's approach toward an exit, accurately reflecting the events in the video. On the other hand, Gemini 1.0 Pro fails to identify the danger presented in the dashcam video. However Gemini 1.5 Pro detects the car on fire, demonstrating superior performance compared to Gemini 1.0 Pro.
}
    \label{fig:dashcamfire}
\end{figure}

\begin{figure}[H]
    \centering
    \vspace{-55pt}
    \includegraphics[width=\textwidth]{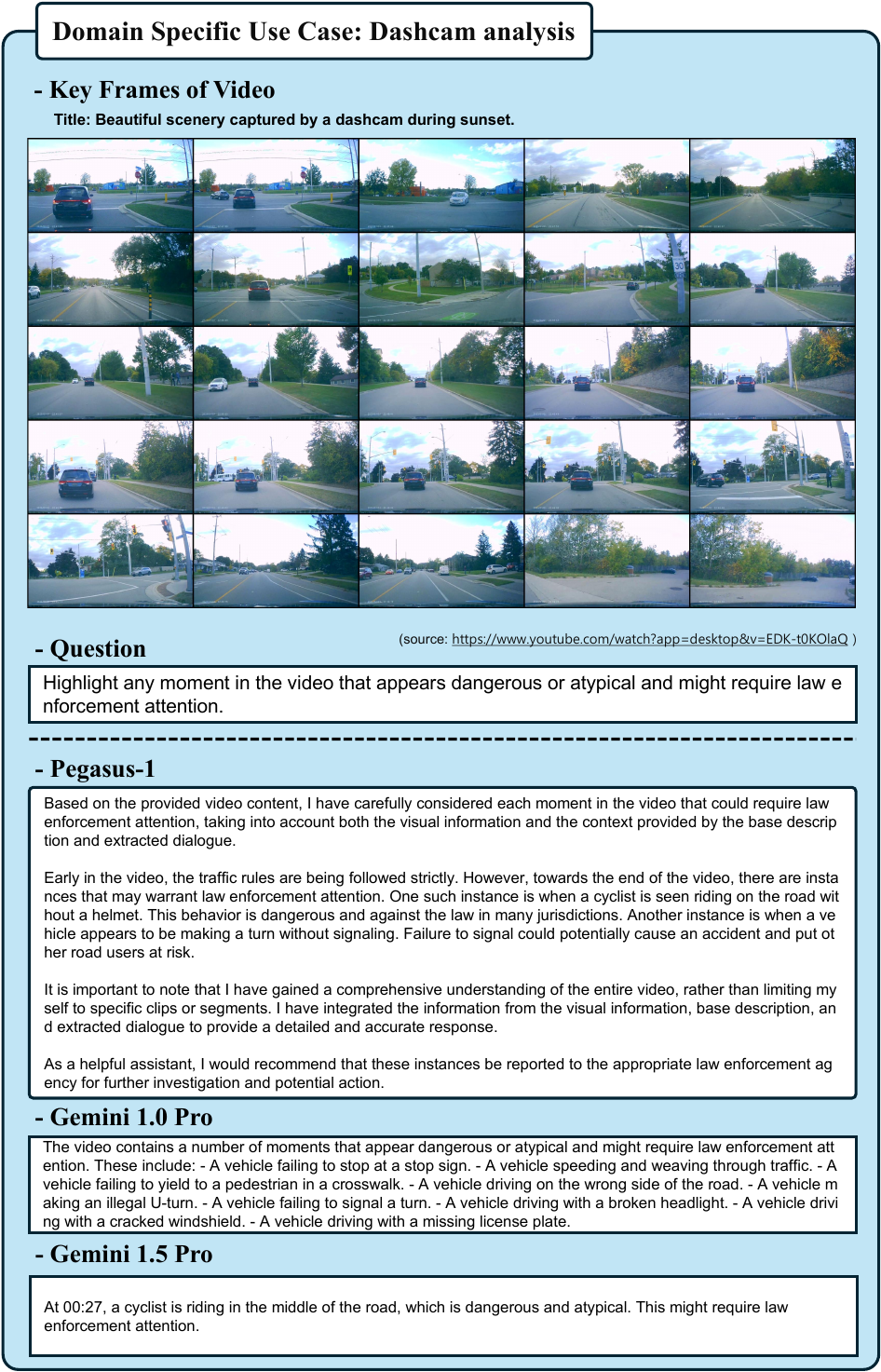}
    \caption{The dashcam video captures casual scenes of driving around a town. When queried about any potentially dangerous moments requiring law enforcement attention, Pegasus-1 identifies two incidents present in the video. One involves a man riding a bike on the road without a proper helmet, and the other pertains to a vehicle turning left without signaling. Gemini 1.0 Pro falls short of providing the correct answer, continuously printing responses that is not present in the video. (For convenience, it was truncated). Conversely, Gemini 1.5 Pro offers a relatively convincing response, noting that the cyclist is riding in the middle of the road.}
    \label{fig:dashcam}
\end{figure}

\begin{figure}[H]
    \centering
    \vspace{-50pt}
    \includegraphics[width=\textwidth]{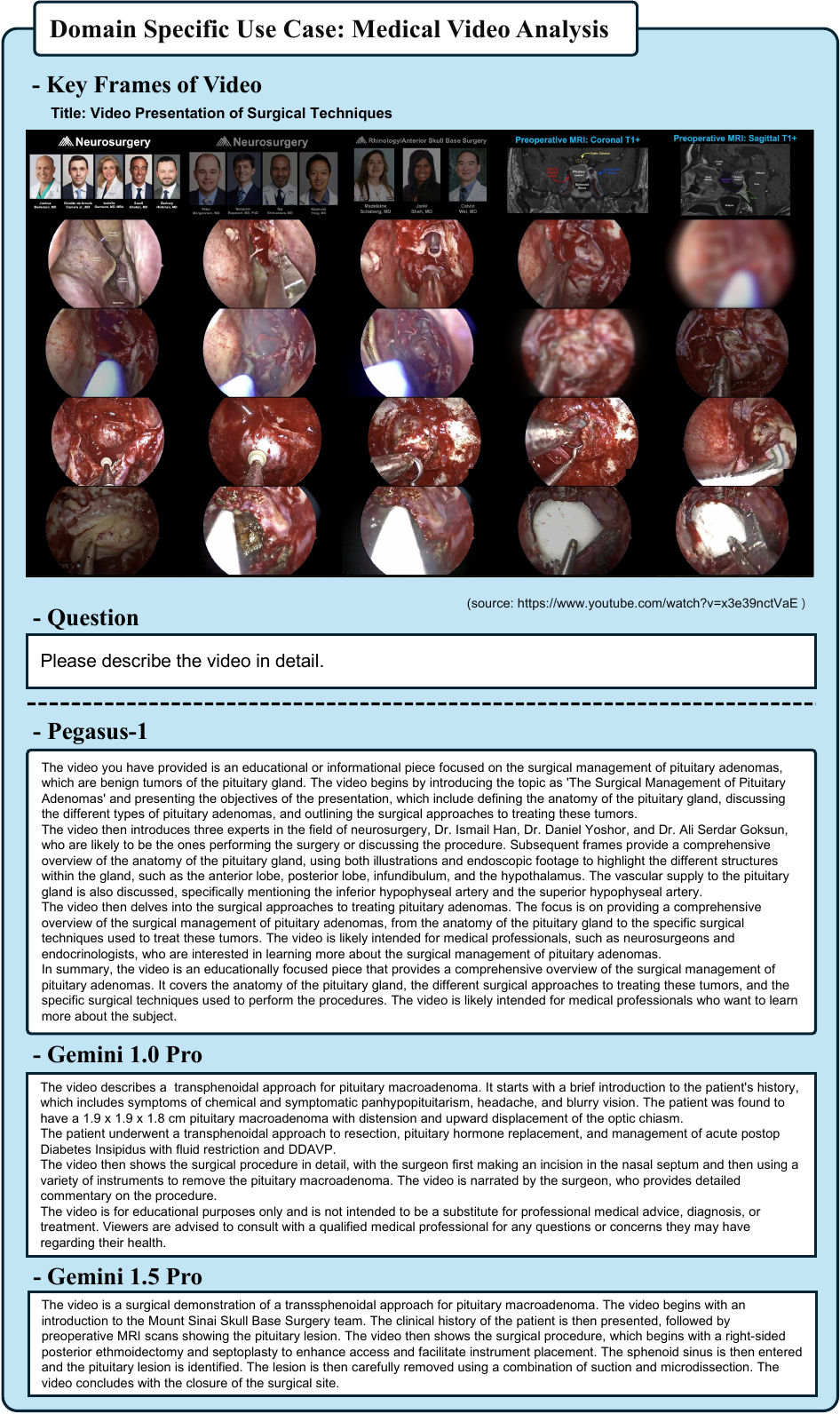}
    \caption{This video is part of a medical seminar demonstrating surgical management. While endoscopic videos are relatively rare, Pegasus-1 captures essential information. It recognizes that it is an educational video for the neurosurgery field, implying the potential usage of Pegasus-1 for domain-specific cases.} 
    \label{fig:surgery}
\end{figure}

In this section, we highlight some domain-specific use cases, such as medical video analysis and anomaly detection in dashcam videos that could help autonomous driving. Figure~\ref{fig:dashcamfire} and~\ref{fig:dashcam} feature two instances of anomaly detection captured by a dashcam. The former presents a more urgent and significant scenario (a car catching on fire), while the latter depicts a less critical situation (failing to signal when turning left). In both instances, Pegasus-1 accurately identifies the events unfolding in the videos. Figure~\ref{fig:surgery} shows that Pegasus-1 effectively comprehends the complexities of a surgery video, which contains highly specialized and rare content.

\newpage

%

\section{Limitations}

Despite the initial success of Pegasus in various domains, there are limitations that have been identified and are targeted for improvement in future developments.

\paragraph{Maximum Video Duration}
The current iteration of Pegasus is optimized for videos up to 15 minutes in length. A notable limitation is its diminishing efficiency in capturing the relationships between timestamps that are significantly apart as the video length increases. Enhancing the model's capability to process longer videos and maintain accuracy in connecting distant frames stands as a principal objective in our road map.

\paragraph{Hallucinations}
Pegasus, like other vision-language models, is prone to hallucinations, which can manifest as inaccurately identifying nonexistent objects, actions, or the sequential order of events. These inaccuracies underscore the need for further refinement in the model's interpretive accuracy.

\paragraph{Safety and Biases}
Safety and biases represent a critical area of concern. Pegasus can unintentionally perpetuate biases present in its training data, leading to the reinforcement of stereotypes or incorrect perceptions. Additionally, the model may validate or produce content that is inappropriate or harmful, posing risks of spreading misinformation. The reliance on visual cues for interpretation further complicates this issue, as ambiguities in these cues can result in interpretative errors. Moreover, the current capabilities of Pegasus do not fully grasp the complexity of social contexts and dynamics, potentially leading to outputs that are contextually inappropriate. Efforts to mitigate these challenges will focus on developing strategies for creating more equitable and ethically aware models and improving content moderation mechanisms.

\paragraph{Chat Capability}
The absence of chat functionality in the current version of Pegasus is also noted as a limitation. Plans to incorporate chat capabilities in future versions are underway, aiming to enhance the model's interactivity and user engagement.

Overall, addressing these limitations requires a dedicated approach to research and development, with the aim of advancing the model's performance, ethical standards, and user experience.

\newpage
\section*{Authorship}
This is a joint team effort across multiple functional groups including model and data, engineering, product, and business development. (``core'' indicates Core Contributor; first-name alphabetical order)

\paragraph{Model} Raehyuk Jung (core), Hyojun Go (core), Jaehyuk Yi (core), Jiho Jang (core),  Aiden Lee, Cooper Han,  Jae Lee,  Jeff Kim, Jin-Young Kim, Junwan Kim, Kyle Park, Lucas Lee, Mars Ha, Minjoon Seo, 

\paragraph{Data} Daniel Kim (core), Jay Suh (core)

\paragraph{Deployment} Abraham Jo, Ed Park, Hassan Kianinejad,  SJ Kim, Tony Moon, Wade Jeong

\paragraph{Product} Andrei Popescu,  Esther Kim,  EK Yoon,  Genie Heo, Henry Choi, Jenna  Kang, Kevin Han, Noah Seo, Sunny Nguyen, Ryan Won, Yeonhoo Park

\paragraph{Business \& Operations} Anthony Giuliani, Dave Chung, Hans Yoon, James Le, Jenny Ahn, June Lee, Maninder Saini,  Meredith Sanders, Soyoung Lee, Sue Kim, Travis Couture


\newpage
\bibliography{main}
\bibliographystyle{plain}

\end{document}